\documentclass[10pt, twocolumn, twoside]{IEEEtran}
\usepackage{graphicx}
\usepackage{amsmath}
\usepackage{amsthm}
\usepackage{amsfonts}
\usepackage{caption}
\usepackage{subcaption}
\usepackage{siunitx}
\usepackage{amssymb}
\usepackage{placeins}
\usepackage{lettrine}
\usepackage{multicol, blindtext}
\usepackage{bbm}
\usepackage{color}
\usepackage{algorithm}
\usepackage{algpseudocode}
\usepackage{authblk}
\usepackage{lipsum}
\usepackage[strings]{underscore}

\newcommand\blfootnote[1]{%
  \begingroup
  \renewcommand\thefootnote{}\footnote{#1}%
  \addtocounter{footnote}{-1}%
  \endgroup
}
\newtheorem{remark}{Remark}
\makeatletter

\newcommand{\Rmnum}[1]{\expandafter\@slowromancap\romannumeral #1@}
\newtheorem{theorem}{Theorem}
\newtheorem{lemma}{Lemma}
\newtheorem{definition}{Definition}
\makeatletter
\def\BState{\State\hskip-\ALG@thistlm}
\makeatother
\ifCLASSINFOpdf
\else
\fi

\begin{document}
%
\title{On Optimal Scheduling for Joint Spatial Division and Multiplexing Approach in FDD Massive MIMO}
\author[*]{Ali Maatouk}
\author[*]{Salah Eddine Hajri}
\author[*]{Mohamad Assaad}
\author[$\S\dagger$]{Hikmet Sari}
\affil[*]{TCL Chair on 5G, Laboratoire des Signaux et Syst\`emes, CentraleSup\'elec, Gif-sur-Yvette, France }
\affil[$\S$]{NUPT, 66 Xinmofan Road, Gulou District, Nanjing, 210003 China}
\affil[$\dagger$]{Sequans Communications, 15 -- 55 Boulevard Charles de Gaulle, 92700 Colombes, France}
\maketitle

\blfootnote{This work has been performed in the framework of the Horizon 2020 project ONE5G (ICT-760809) receiving funds from the European Union.}
\blfootnote{The material in this paper is an extention of the work originally presented in IEEE ICC 2018 \cite{2017arXiv171203022M}.}
\begin{abstract}
Massive MIMO is widely considered as a key enabler of the next generation 5G networks. With a large number of antennas at the Base Station, both spectral and energy 
efficiencies can be enhanced. Unfortunately, the downlink channel estimation overhead scales linearly with the number of antennas. This burden is easily mitigated in 
TDD systems by the use of the channel reciprocity property. However, this is unfeasible for FDD systems and the method of two-stage beamforming was therefore developed to 
reduce the amount of channel state information feedback. The performance of this scheme being highly dependent on the users grouping and scheduling 
mechanims, we introduce in this paper a new similarity measure coupled with a novel clustering procedure to achieve the appropriate users grouping. We also proceed to formulate the optimal users scheduling policy in JSDM and prove that it is NP-hard. This result is of paramount importance 
since it suggests that, unless P=NP, there are no polynomial time algorithms that solve the 
general scheduling problem to global optimality and the use of sub-optimal scheduling strategies is more realistic in practice. We therefore use graph theory to 
develop 
a sub-optimal users scheduling scheme that runs in polynomial time and outperforms the scheduling schemes previously introduced in the literature for JSDM 
in both sum-rate and throughput fairness.
\end{abstract}



%
\IEEEpeerreviewmaketitle

\section{Introduction}
\lettrine{M}{obile} traffic demand has never been as high as it is today due to the widespread of smart-phones and the rise of data-hungry applications like video 
streaming. The next generation mobile networks should, therefore, be able to keep up with the high throughput demand. Massive Multiple-Input and Multiple-Output (\textbf{MIMO}) \cite{5595728} is considered one of 
the promising technologies that will enable the next generation mobile networks to cope with this demand. In comparison to the current multi-user MIMO systems, massive MIMO incorporates a significantly higher number of antennas at the Base Station. This has been shown to offer superior performance in terms of both energy efficiency and overall capacity 
\cite{6457363} which made massive MIMO a hot research topic and a key component of future standards \cite{wong_schober_ng_wang_2017}.

Although the original
Massive MIMO concept \cite{5595728} assumes Time Division Duplex (\textbf{TDD}), the study of massive MIMO for Frequency Division Duplex (\textbf{FDD}) systems is of paramount importance as they still represent the vast majority of 
the currently deployed cellular networks. On top of that, FDD systems exhibit a better performance in scenarios with symmetric traffic and
delay sensitive applications \cite{2017arXiv170804444B}. However, the high number of antennas will result in complications in terms of downlink channel estimation and feedback for FDD systems. This comes from the fact that the downlink channel estimation overhead scales linearly with the number of antennas \cite{6375940}. This is mitigated in TDD systems by exploiting the channel reciprocity since the channel estimate of the 
uplink direction can be directly utilized for the downlink direction \cite{6375940}\cite{7852217} which is not feasible in FDD systems.\color{black}

%

In order to deal with this difficuly, Joint Spatial Division and Multiplexing (\textbf{JSDM}), an approach to multiuser MIMO downlink that is considered one of the 
most promising candidates for FDD massive MIMO, was proposed by the authors in \cite{6542746}. The idea revolves around partitioning users with the same channel Second 
Order Statistics (\textbf{SOS}) into groups and splitting the downlink beamforming precoder into two stages: an outer precoder, that solely depends on the channel SOS, and an 
inner precoder that depends on the instantaneous realization of the \emph{effective} channel with the dimensions of the \emph{effective} channel being significantly 
less than the number of antennas, thanks to the outer precoder projection. The authors in \cite{6542746} were able to show that even with reduced Channel State Information at the Transmitter (\textbf{CSI}), JSDM achieves 
the same sum capacity of the corresponding MU-MIMO broadcast channel if the eigenspaces of groups are mutually orthogonal, a condition that was given the name "tall 
unitary".

In realistic scenarios, users may have similar but not necessarily identical channel SOS. This dictates us to incorporate a clustering process that finds the 
appropriate partitioning of the users into groups with \emph{sufficiently} similar covariance eigenspaces.
 Another thing to point out is that
 with a high number of 
users uniformly distributed across the cell, the eigenspaces of the groups are far from meeting the tall unitary condition. This forces us to reduce the number of 
simultaneously served groups by the use of a smart scheduling scheme. These issues inspired the work in \cite{6778043} where $K$-means clustering process was adopted 
and a greedy sum-rate maximization scheduling algorithm was proposed. 

The fact that the scheduling scheme proposed in \cite{6778043} is greedy in nature, and in the aim of simplifying the users partitioning process, recent work \cite{7997313} adopted 
a hierarchical clustering algorithm which mixes both target number of clusters and chordal distance threshold to reach an appropriate users clustering. A scheduling 
scheme that is based on improving the average Signal to Leakage plus Noise Ratio (\textbf{SLNR}) of the system was also proposed and was shown to outperform in terms of sum-rate all the previous methods in the literature of JSDM \cite{7997313}.

Our paper aims to deal with the issues that are still present in the previous approaches. To that end, the following are the key contributions of this paper:
\begin{itemize}
\item First, the optimum number of clusters in the network is not known beforehand and choosing a random number of clusters can have severe impact on the performance of JSDM. On top of that, any thresholds involved in the clustering process is hard 
to predict when using the standard chordal distance as a similarity measure. The first contribution of the paper consists of adopting a novel similarity measure along with a new clustering scheme where the number of clusters is not required to be given as an input. 
\item Secondly, knowing that the eigenspaces of groups are far from being orthogonal, one may seek to deal with 
the inter-group interference by applying appropriate outerprecoding techniques. Since this approach is unable to completely eliminate interference in realistic scenarios as will be shown in the paper, adopting a scheduling scheme to reduce the number of served groups becomes of \emph{paramount importance}. One may also use the outer precoder to match each group's covariance eigenspace in the aim of getting the highest useful signal while dealing with the inter-group interference by solely relying on 
the scheduling process. A numerical comparison between these two approaches is presented in our paper which will lead to an interesting conclusion that in certain scenarios, 
inter-group interference is better to be dealt with solely on the MAC layer.
\item With the scheduling process being pivotal for both outerprecoder approaches, a smart scheduling scheme is 
to be adopted to extract the best possible performance of JSDM.
 Due to the special structure of the achievable rate in JSDM, as will be seen in the sequel, the conventional scheduling techniques for wireless networks fail and new scheduling propositions have been introduced in the literature for JSDM (e.g. \cite{6778043}\cite{7997313}). Although there have been many scheduling methods propositions in the JSDM literature such as \cite{6778043}\cite{7997313} to 
maximize the sum-rate, none have previously investigated the complexity of the optimal users scheduling policy for JSDM. The \emph{main technical} contribution of the paper is the 
establishment of the NP-hardness of finding the optimal users scheduling solution in JSDM. The scheduling problem is formulated as a weighted sum-rate
subject to certain constraints and the proof of NP-Hardness is provided. The proof is novel and original as it relies on a specific decomposition of users' groups and several mathematical lemmas to provide the justification of the polynomial reduction from the SAT problem, a well known NP-Complete problem, to the proposed scheduling problem. Moreover, these complexity
results are not bound to a particular clustering process and
therefore hold for any users clustering techniques used.
\item As suggested by our complexity results, finding
the optimal scheduling policy for any given JSDM problem is generally intractable. Thus,
instead of insisting on finding an efficient algorithm that is able
to find the global optimum, one has to settle for polynomial
time sub-optimal schemes that perform well. Our final contribution revolves around developing a well performing scheduling scheme for JSDM that runs in polynomial time. To do so, we take an interesting 
approach of modeling our problem using graph theory tools. Afterwards, by applying appropriate transformations on our graph, we are able to use well-known approximations algorithms from the vast graph theory literature. The result is a polynomial time 
running scheduling scheme that outperforms all previously proposed scheduling methods for JSDM in both sum-rate and throughput fairness.
\end{itemize}
The rest of the paper is organized as follows: Section \Rmnum{2} presents the system model. Section \Rmnum{3} introduces the newly proposed metric and clustering process. Section \Rmnum{4} includes a discussion on the outer precoder design and the development of our scheduling scheme. Section \Rmnum{5} provides the numerical results while Section \Rmnum{6} concludes the paper.\color{black}
\section{System Model}
We consider a single cell downlink multi-user MIMO system with $N_t$ antennas at the BS and $K$ single-antenna users. The received vector by the users $\boldsymbol{y} \in \mathbb{C}^{K\times 1}$ can be expressed as: 
\begin{equation}
\boldsymbol{y}=\boldsymbol{H}^H\boldsymbol{x}+\boldsymbol{z}
\end{equation}
where $\boldsymbol{x} \in \mathbb{C}^{N_t\times 1}$ is the transmitted signal vector, $\boldsymbol{z} \sim \mathcal{CN}(0,\boldsymbol{I}_K) \in \mathbb{C}^{K\times 1}$ denotes the AWGN vector and $\boldsymbol{H} \in \mathbb{C}^{N_t\times K}$ is the channel matrix. The transmitted signal vector is actually a precoded version of the data vector, i.e, $\boldsymbol{x}=\boldsymbol{V}\boldsymbol{d}$ where $\boldsymbol{V} \in \mathbb{C}^{N_t\times S}$ is the precoder and $\boldsymbol{d} \in \mathbb{C}^{S\times 1}$ is the data vector. The dimension S denotes the total number of independent streams and is upperbounded by $min\{N_t,K\}$\cite{6542746}. As previously adopted in \cite{6542746} and for the sake of simplicity, we adopt the approach of equal power allocation (EPA) i.e. $\mathbb{E}(\boldsymbol{d}\boldsymbol{d}^H)=\frac{P}{S}\boldsymbol{I}_S$ where $P$ is the total downlink power budget. As for the channel model, we adopt a Rayleigh fading channel (i.e. no Line-Of-Sight propagation) and therefore $\boldsymbol{h}_k\sim\mathcal{CN}(0,\boldsymbol{R}_k)$ where $\boldsymbol{R}_k$ is a positive semi-definite covariance 
matrix. By taking the Eigen Value Decomposition (\textbf{EVD}) of $\boldsymbol{R}_k$, we have the following: 
\begin{equation}
\boldsymbol{R}_k=\boldsymbol{U}_k\boldsymbol{\Lambda}_k\boldsymbol{U}_k^H
\end{equation} 
where $\boldsymbol{\Lambda}_k$ is an $r_k\times r_k$ diagonal matrix with the $r_k$ eigenvalues as diagonal entries. Therefore, $r_k$ represents the rank of $
\boldsymbol{R}_k$ and $\boldsymbol{U}_k \in \mathbb{C}^{N_t\times r_k}$ is nothing but the set of eigenvectors corresponding to the non-zero eigenvalues. By taking the similarity of their channel covariance into account, users are partitioned into G groups with each containing $K_g$ users such as $K=\sum_{g=1}^{G}K_g$. After the 
partitioning, a single representative of the covariance space for the whole group is taken and groups are therefore treated as a single entity. 
JSDM \cite{6542746} revolves around the idea of splitting the precoder $\boldsymbol{V}$ into two stages: $\boldsymbol{V}=\boldsymbol{B}\boldsymbol{P}$, where $
\boldsymbol{B}$ and $\boldsymbol{P}$ are referred to as the outer and inner precoders respectively. The outer precoder $\boldsymbol{B}$, of dimensions $N_t\times b$, is 
based on the channel statistics which is supposed to be known at the base station as adopted in \cite{6542746}\footnote{The channel statistics vary at a much slower 
rate than the channel coherence time and therefore can be assumed to be locally stationary and easily tracked by methods cited in \cite{6542746}}. The design of the outer precoder 
can be aimed to minimizing \emph{inter}-group interference. On the other hand, one may use the outer precoder to match the covariance space of scheduled groups in order to obtain the highest useful signal possible. The \emph{inter}-group interference is therefore left to be dealt with only on the MAC-Layer (i.e. 
by appropriate scheduling groups). The design of the outer precoder 
will be further detailed in Section \Rmnum{4}-A. As for the inner precoder $\boldsymbol{P}$, it is of dimensions $b\times S$ and depends on the instantaneous channel realizations and is intended to suppress \emph{intra}-group interference.
By considering the partitioning of the users, we have the following: $\boldsymbol{H}_g=[\boldsymbol{h}_{g_1},\ldots,\boldsymbol{h}_{g_{K_g}}]$, $\boldsymbol{H}=[\boldsymbol{H}_1,\ldots,\boldsymbol{H}_G]$, $\boldsymbol{B}=[\boldsymbol{B}_1,\ldots \boldsymbol{B}_G]$, $\boldsymbol{P}=diag\{\boldsymbol{P}_1,\ldots,\boldsymbol{P}_G\}$ and we define the effective channel $\widetilde{\boldsymbol{H}}=\boldsymbol{B}^H\boldsymbol{H}$.
It is straightforward that the effective channel is of dimension $b\times K$ with $b=\sum_{g=1}^{G}b_g$ and $b_g\ll N_t$. In fact, the drastic reduction in the amount of CSI feedback takes place when each user $g_k$ has to feedback his \emph{effective} channel $\widetilde{\boldsymbol{h}}_
{g_k} \in \mathbb{C}^{b_g\times 1}$ rather than $\boldsymbol{h}_{g_k} \in \mathbb{C}^{N_t\times 1}$. We will refer to 
this approach as \emph{Per Group Processing} (\textbf{PGP}).\\
The received signal by group g can be therefore written as: 
\begin{equation}
\boldsymbol{y}_g=\boldsymbol{H}_g^H\boldsymbol{B}_{g}\boldsymbol{P}_g\boldsymbol{d}_g +\sum_{g'\neq g}{\boldsymbol{H}}_g^H\boldsymbol{B}_{g'}\boldsymbol{P}_{g'}\boldsymbol{d}_{g'}+\boldsymbol{z}_g
\label{received}
\end{equation}
where $\boldsymbol{d}_g \in \mathbb{C}^{S_g\times 1}$ with $S_g$ being the number of independent streams intended for group $g$.
By adopting the PGP approach and assuming perfect \emph{effective} CSI at the BS, a Zero Forcing (\textbf{ZF}) inner precoder can be calculated as follows:
\begin{equation}
\boldsymbol{P}_{g}=\zeta_g\widetilde{\boldsymbol{H}}_g(\widetilde{\boldsymbol{H}}^H_g\widetilde{\boldsymbol{H}}_g)^{-1} \in \mathbb{C}^{b_g\times S_g}
\end{equation}
with $\zeta_g$ being a normalization constant to ensure that the power budget constraint is satisfied:
\begin{equation}
\zeta_g^2=\frac{S_g}{tr(\boldsymbol{B}_g\widetilde{\boldsymbol{H}}_g\bigg(\widetilde{\boldsymbol{H}}^H_g\widetilde{\boldsymbol{H}}_g{\bigg)}^{-2}\widetilde{\boldsymbol{H}}^H_g\boldsymbol{B}^H_g)}
\end{equation}

\section{Correlation Clustering}
In order to effectively exploit the JSDM approach, users in the cell must be divided into groups in a way that users within each group have similar channel covariance. The necessity of this criterion comes from the fact that outer precoding techniques treat each group as a single entity and therefore assigning users with distant covariances to the same group will result in a huge performance degradation. Knowing that users might have \emph{similar} but not necessarily identical covariance matrices, the appropriate grouping of the users is therefore vital for the work. The research papers that investigated this clustering problem presented two approaches: $K$-means clustering \cite{6778043} and a hierarchical clustering \cite{7997313}.
Both of these approaches used the chordal distance as a similarity metric. The disadvantage of such a metric comes from the fact that the prediction of any threshold involved in the clustering process is a difficult task. Motivated by this, we adopt in the following subsection a new similarity measure suitable for our problem.
\vspace{-22pt}
\subsection{Similarity Measure}
A novel correlation distance metric was firstly introduced by Herdin et al. in their paper \cite{1543265} and was given the name Correlation Matrix Distance (\textbf
{CMD}). It was used to track the changes of spatial structures of the channel in non-stationary MIMO. It was not long before the use of this metric was extended to 
many different research work. For instance, the authors in \cite{7166317} used it in the context of Grassmannian subspace packing. The same metric was also adopted by 
the authors in \cite{6135456} to study the effect of subspace alignment in multi-user MIMO. 
In the previous literature that investigated the grouping process \cite{6778043}\cite{7997313}, the covariance similarity between two users $1$ and $2$ was solely 
taken based on their covariance's eigenstructures $(\boldsymbol{U}_1\boldsymbol{U}_1^H,\boldsymbol{U}_2\boldsymbol{U}_2^H)$ without taking into account the energy of 
the modes. In this paper, we will be applying our similarity measure on the whole covariance matrices $(\boldsymbol{R}_1,\boldsymbol{R}_2)$. The motivation behind this 
is that differences in the eigenstructures of weak modes should contribute less than the ones of strong modes. For instance, consider the case where the covariance 
space of two users differs only in the low energy modes. The similarity between these two users should still remain high which is not the case if we solely take into 
account the covariance's eigenstructures. Based on CMD, we can define the new similarity measure as follows:
\begin{equation}
d_s(\boldsymbol{R}_1,\boldsymbol{R}_2)=1-CMD(\boldsymbol{R}_1,\boldsymbol{R}_2)=\frac{Tr(\boldsymbol{R}_1^{H}\boldsymbol{R}_2)}{||\boldsymbol{R}_1||_F.||\boldsymbol{R}_2||_F}
\end{equation}
One can clearly see that our measure is lower bounded by $0$ and upper bounded by $1$. The lower bound corresponds to the case where $\boldsymbol{R}_1$ and $\boldsymbol
{R} _2$ are orthogonal while the similarity reaches its upper bound when $\boldsymbol{R}_1$ and $\boldsymbol{R}_2$ are collinear. This proposed measure can be therefore 
regarded as an extension of the well-known cosine similarity of vectors ( a widely used metric in clustering schemes see, e.g., \cite{6933489}) to matrices 
and therefore can now be considered as what we will call \emph{Degree of OverLap} (\textbf{DOL}) between the two covariance spaces. To the knowledge of the authors, this is the first time it 
has been used in the context of users clustering for FDD massive MIMO. The advantages of this proposed similarity measure in comparison to the chordal distance counterpart can be summarized in the following:
\begin{enumerate}
\item The proposed similarity measure is normalized, which makes it more sensitive to differences in the correlation structure of the users \cite{6135456}
\item The proposed similarity measure is upperbounded by $1$ and lowerbounded by $0$, i.e. the search space for any desired clustering threshold is small which is an appealing property as will be detailed in Section V-C
\end{enumerate}\color{black}
\subsection{Clustering Algorithm}
Unlike the previously proposed schemes, we aim to employ a clustering algorithm that does not have the target number of clusters as an input. To do so, we take advantage of the ease of threshold design presented by our proposed similarity metric.
An interesting way to do so is to choose $DOL_{th}$ high enough such as if $d_s(\boldsymbol{R}_k,\boldsymbol{R}_{k'})\geq DOL_{th}$ then users $k$ and $k'$ can be considered to be laying in the same correlation space. Unlike other metrics, this threshold is easily determined. In fact, one can simply say if the degree of overlap between the two spaces is above 0.8-0.9 then consider them as highly similar and are preferred to be assigned to the same cluster (see Section \Rmnum{5}-C for an in-depth discussion). Based on this, we can construct what we will call a \emph{complete advice graph} $G_c=(V_c,E_c)$. In this graph, each vertex represents a user and an edge $e \in E_c^+$ would have a $\langle
+1\rangle$ label to signal that these two users are advised to be in the same cluster while any edge $e \in E_c^-$ would have a $\langle
-1\rangle$ label to refer to the opposite case. What makes this modeling interesting is that by seeking clusters made of vertices with positive edges between them, we are sure that the criterion of similar covariance in each group previously mentioned will be met. Our goal therefore becomes to produce a partition of the graph's vertices in a way that agrees as much as possible with the edge labels. To do so, we make use of the vast graph theory literature, more particularly the \emph{Correlation Clustering} literature \cite{Bansal2004}. The correlation clustering seeks a partition of the graph's vertices based on minimizing a cost function $J$ referred to as the total disagreements. The total disagreements of the resulting partitioned graph is defined as the overall negative weights inside a cluster added to the positive weights between clusters. 
Our partitioning problem can be formulated as follows:
\begin{equation}
\begin{aligned}
& \underset{x_{uv} \forall (u,v) }{\text{minimize}}
& & J=\sum_{(u,v)\in E_c^+}x_{uv}+\sum_{(u,v)\in E_c^-}(1-x_{uv})\\
& \text{subject to}
& & x_{uv}+x_{vw}\geqslant x_{uw} \:\:\forall u,v,w \in V_c\\
&&& x_{uv}=x_{vu} \:\:\forall u,v\in V_c
\end{aligned}
\label{LP}
\end{equation}
where $x_{uv}$ is a binary variable which is null when $(u,v)\in V_c\times V_c$ are assigned to the same cluster and is $1$ otherwise. 
The constraints found in (\ref{LP}) account for the symmetry of $x_{uv}$ and the triangular inequality\footnote{The triangular inequality ensures that our solution 
respects the fact that if $(u,v)$ are assigned to the same cluster and $(v,w)$ are assigned to the same cluster then $(u,w)$ 
should be in the same cluster as well} satisfied by these binary variables. The interesting aspect of this partitioning formulation is that there is no need to 
give the target number of clusters as input. In fact, the resulting optimal number of clusters could be any value
from 1 to K depending on what fits our graph the most. In general, solving (\ref{LP}) and finding the optimal clustering is NP-hard, as proven in \cite{Bansal2004} 
using a reduction from the Exact Cover by 3-Sets (\textbf{X3C}) problem which is one of Karp's 21 \emph{NP-complete} problems. To deal with this complexity, one can turn the 
problem into a Linear Program (\textbf{LP}) simply by relaxing the binary constraint and substituting it with $x_{uv}\in [0,1] \:\:\forall u,v\in V_c$. The LP is then 
solved in polynomial time by any desired standard LP solvers followed by appropriate rounding of the fractional values. The question that arises: how to round the 
fractional solutions of the LP? The literature is rich with rounding techniques that achieve decent performance guarantees. The most recent work \cite{Chawla:2015:NOL:2746539.2746604} revolves around a randomized technique named \emph{Pivoting}.
 Details concerning this procedure are presented in Appendix A-A and we provide in Algorithm 1, a summary of the clustering scheme.
\begin{algorithm}
\caption{Clustering Scheme}\label{euclid}
\begin{algorithmic}[1]
\State \textbf{Init.} Compute the similarity matrix $\boldsymbol{S}_{ij}=d_{s}(\boldsymbol{R}_i,\boldsymbol{R}_j)$
\If {$s_{ij} > DOL_{th}$} $s_{ij}=+1$
\Else \:$s_{ij}=-1$
\EndIf
\State Solve (\ref{LP}) to get $x_{uv}$ then apply (\ref{funct}) to get $p_{uv}=f(x_{uv})$
\Procedure{Pivoting}{}
\State Let $V_0=V_c$ the set of all vertices, let $t=0$
\While {$V_t\neq \emptyset$}
\State Pick a pivot $w_t\in V_t$ randomly and let $S_t=w_t$
\State $\forall\:u\in V_t$, add $u$ to $S_t$ with probability $1-p_{wu}$
\State let $V_{t+1}=V_{t} \setminus S_t$, let $t=t+1$
\EndWhile
\EndProcedure
\State Output the clusters $S_{0},\ldots,S_{Final}$
\end{algorithmic}
\end{algorithm}
\section{Downlink Scheduling}
After grouping users with similar second order channel statistics, we can now deal with the orthogonality aspect of JSDM. In realistic scenarios, groups do not lay in 
mutual orthogonal channel covariance spaces and inter-group interference can therefore limit the overall performance. One can seek to reduce this interference by 
applying appropriate outer precoding techniques but it is insufficient as will be proven shortly. Therefore, adopting a scheduling scheme to deal with the \emph
{residual} interference is of paramount importance for the overall performance of JSDM. One can also follow a different approach: use the outer precoder techniques to get the highest useful signal possible and deal with the inter-group interference itself \emph{solely} on the MAC layer. The differences between these two approaches and the proposed scheduling scheme are both presented in this section.
\vspace{-12pt}
\subsection{Outer Precoder}
\begin{figure}[!ht]
\centering
\includegraphics[width=.4\linewidth]{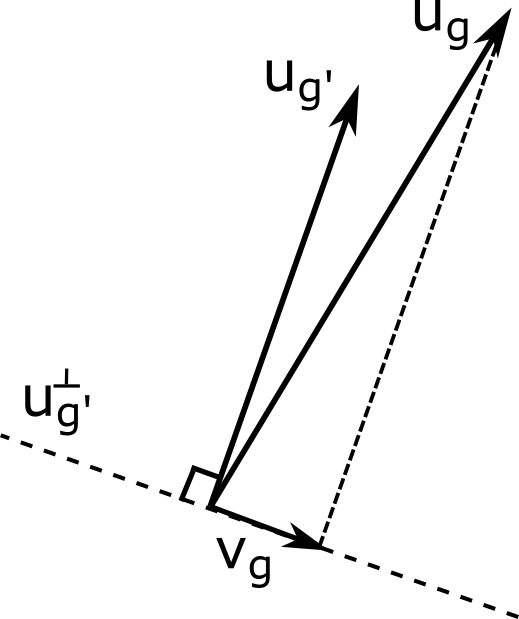}
\setlength{\belowcaptionskip}{-10pt}
\caption{Projection}
\label{Projection}
\end{figure}

We define the group's covariance centroid that would be taken as a representative of each group's equivalent covariance as:
\begin{equation}
\boldsymbol{R}_g=\frac{1}{K_g}\sum_{k=1}^{K_{g}}\boldsymbol{R}_{g_k}\overset{\textbf{EVD}}{=}\boldsymbol{U}_g\boldsymbol{\Lambda}_g\boldsymbol{U}_g^H \:\: \text{with} \:\: \boldsymbol{U}_g \in \mathbb{C}^{N_t\times r_g}
\end{equation}
where $r_g$ is the rank of the centroid. The advantage of our clustering scheme is highlighted here. By choosing a \emph{high} $DOL_{th}$, we know that the covariances of users in each group will be really similar and therefore the centroid is a 
good representative of each cluster. In fact, the threshold is set in a way to ensure having a good representation of each cluster's equivalent covariance. However this is not necesarrily achieved by employing other clustering algorithms with \emph{pre-determined target number of clusters} as in \cite{6778043}\cite{7997313} since it is hard to predict beforehand the exact number of clusters for which each group's centroid is a good representative. As for the outer precoder design, one can follow either of these two approaches:\\
\textbf{Approach 1:} This approach seeks to eliminate inter-group interference at the physical layer level. As proposed in \cite{6542746} and adopted in \cite{7997313}, one can do so by 
employing appropriate outerprecoding techniques and was given the name \emph{Approximate Block Diagonalization}. This is done by first building the inter-group 
interference matrix for group $g$ and projecting the intended signal space on the interference matrix's orthogonal space. The interference matrix as seen by group $g$ 
is $\boldsymbol{\Xi}_g=[\boldsymbol{U}_1^*,\ldots,\boldsymbol{U}^*_{g-1},\boldsymbol{U}_{g+1}^*,\ldots,\boldsymbol{U}_G^*]$ which is composed of the eigenspaces of 
other \emph{active} groups with $\boldsymbol{U}_{g'}^* \in \mathbb{C}^{N_t\times r_{g'}^*}$ where $r_{g'}^*$ is a design parameter that represents the number of high 
energy channel modes\footnote{This is also one of the motivation why the energy of the channel modes was considered in our similary measure since the strongest modes 
are taken into account first in the outer precoder design} taken into account. By considering the Singular Value Decomposition (\textbf{SVD}) of $\boldsymbol{\Xi}_g$, we can decompose  the 
set of left eigenvectors as $[\boldsymbol{E}_g^{(1)},\boldsymbol{E}_g^{(0)}]$ where $\boldsymbol{E}_g^{(0)}$ is of dimension $N_t\times (N_t-\sum_{g'\neq g}r_{g'}^*)$ 
and forms a unitary basis for $Span^{\bot}({\boldsymbol{U}_{g'}^* : g'\neq g})$. Based on this, we can construct our projected channel covariance matrix as follows:
\begin{equation}
\widehat{\boldsymbol{R}}_g=(\boldsymbol{E}_g^{(0)})^H\boldsymbol{U}_g\boldsymbol{\Lambda}_g\boldsymbol{U}_g^H\boldsymbol{E}_g^{(0)}\overset{\textbf{EVD}}{=}\boldsymbol{G}_g\boldsymbol{\Phi}_g\boldsymbol{G}_g^H
\end{equation}
After projection to $Span^{\bot}({\boldsymbol{U}_{g'}^* : g'\neq g})$, the next step would be to match the $b_g$ strongest eigenmodes 
of our projected channel. By considering the EVD of our projected channel, we can decompose the set of eigenvectors as $\boldsymbol{G}_g=[\boldsymbol{G}_g^{(1)},
\boldsymbol{G}_g^{(0)}]$ where $\boldsymbol{G}_g^{(1)}$ is of dimension $(N_t-\sum_{g'\neq g}r_{g'}^*)\times b_g$. Overall, the outer precoder becomes $\boldsymbol{B}
_g=\boldsymbol{E}_g^{(0)}\boldsymbol{G}_g^{(1)}$. One could argue that by simply choosing our design parameter $r_{g'}^*=r_{g'}$ (i.e. including all modes), we will lay in an inter-group interference-free scenario.
However by construction, the channel effective dimension is $b_g\leq rank(\widehat{\boldsymbol{R}}_g)=min(r_g,N_t-\sum_{g'\neq g}r_{g'}^*)$. Therefore, including more 
modes would actually shrink our dimensionality and lead to a dimensionality bottleneck. Keeping in mind that $S_g\leq b_g$, the dimensionality bottleneck is a serious issue 
since we are obliged to serve a certain number of independent streams to each group. Without loss of generality, we suppose the following:
\begin{itemize}
\item $N_t\geq K$, i.e. it is possible to schedule all users together
\end{itemize}  
The previous condition ensures that there are enough transmitting antennas to schedule all users simultaneously and is satisfied in realistic \emph{massive} MIMO 
scenarios. We first define our assignment vector as a $K$-tuplet $\boldsymbol{x}$ of binary variables $x_{g_k}$ with a value $1$ indicating that user $k$ of group $g$ is scheduled. Based on this, we consider that $S_g=\sum_{k=1}^{K_g}x_{g_k}$, in other words we suppose that when a group is scheduled, each \emph{scheduled} user of that group receives an independent stream. In this case, our goal becomes to keep $b_g\geq \sum_{k=1}^{K_g}x_{g_k}$. We propose to find the interference matrix's design parameter $r_g$ by seeking the largest integer $c\geq 1$, that we will give the name \emph{inclusion factor} to, such that:
\begin{equation}
N_t\geqslant\sum_{g=1}^{G}min\{r_g,c(\sum_{k=1}^{K_g}x_{g_k})\}
\end{equation}
Afterwards, we set $r_{g'}^*=min\{r_{g'},c(\sum_{k=1}^{K_g'}x_{g'_k})\}$ which refers to the number of strongest modes of the \emph{active} groups taken into account. We also let $b_g=min\{r_g,N_t-\sum_{g'\neq g}r_{g'}^*\}$ which is now larger than $\sum_{k=1}^{K_g}x_{g_k}$ by construction. This method essentially tries to find the largest number of strong modes of the scheduled groups to include while respecting the dimensionality bottleneck imposed by the necessity of sending $S_g=\sum_{k=1}^{K_g}x_{g_k}$ streams per scheduled group. Since not necessarily all modes are included ($r_{g'}^*\leq r_g$), inter-group interference would still be inevitable and we therefore include scheduling in our design. To note, we have let $b_g$ to be equal to its maximal allowed value since the higher the number of matched strong modes of the projected channel is, the better the performance experienced by group $g$ is to be expected. \color{black}\\
\textbf{Approach 2:} Although the first outerprecoder approach is the most widely used in the literature, we argue that in realistic scenarios, one may recall other design approaches for better performance. In fact, as previously discussed in Section \Rmnum{1}, the groups covariance spaces are far from being orthogonal. Consequently, they may share a decent number of high energy modes. Therefore, by restricting our transmission to each group to be orthogonal to the other groups covariance spaces, the channel modes gain of this group's projected channel can be arbitrarily \emph{small}. To visualize this, consider two strong modes $(u_g,u_{g'})$ as seen in Fig. \ref{Projection} of groups $g$ and $g'$ respectively with the vector length referring to the mode's gain. By restricting the transmission of group $g$ to be orthogonal to $u_{g'}$, the resulting transmitting direction $v_g$ clearly has a much smaller channel gain. 
\\
One may argue if by adopting Approach 1, we are able to schedule groups $g$ and $g'$ but with a much reduced outcome, then perhaps by scheduling groups $g$ and $g'$ one at a time we would gain an overall better performance. Therefore in this approach, we simply match each group's covariance space: 
\begin{equation}
\boldsymbol{B}_g=\boldsymbol{U}_g
\end{equation}
By doing so, we are extracting the highest possible useful signal of each group but the inter-group interference remains high. This high interference can be avoided by appropriate groups scheduling mechanism. In fact, the same scheduling procedure can be used for both outerprecoder approaches. The scheduling scheme works on the residual inter-group interference from the projected channel and on the inter-group interference itself by combining it with Approaches 1 and 2 respectively. With the incorporation of scheduling being important for both cases, answering the following question is of a paramount importance: \emph{Is it better to deal with inter-group interference at both the PHY/MAC or solely at the MAC level in JSDM?} A numerical comparison between the two approaches will be presented in Section \Rmnum{5}-B.
\subsection{Scheduling Problem}
The first step to construct our network utility is to find an expression of the rate $R_{g_k}$ achieved by each user $k$ in group $g$. We employ here a widely used model where the rate $R_{g_k}$ is given by $R_{g_k}(SINR)=log_2(1+SINR_{g_k})$\cite{6542746}. \\
\emph{Large system regime:} The main focus of our paper lays on the case of multi-user \emph{massive} MIMO where the number of antennas and users $N_t,K\longrightarrow 
+\infty$. For this scenario, Random Matrix Theory (\textbf{RMT}) tools come in handy \cite{7084118}. The authors of JSDM \cite{6542746} made use of the work in \cite
{6172680} to propose a deterministic equivalent for the $SINR$ expression in JSDM. Motivated by the fact that this deterministic equivalence was shown to be accurate 
for realistic values of $(N_t,K)$ \cite{6542746}\cite{6172680}, we take it as a basis of our analysis.
Details concerning these equations can be found in \cite{6542746} which for our case 
reduce to:
\begin{equation}
SINR_{g,k}\overset{N_t,K\rightarrow+\infty}{\longrightarrow} \frac{\frac{P}{S}x_{g_k}\overline{\zeta}_g^2}{\sum_{g'\neq g}\sum_{k'=1}^{K_{g'}}\frac{P}{S}x_{g_{k'}'}\overline{\zeta}_{g'}^2\overline{\Upsilon}_{g,g'}+1}
\label{SINR}
\end{equation}
where $x_{g_k}$ is a binary variable that denotes if user $g_k$ is scheduled. $\overline{\zeta}_g^2=\overline{m}_gb_g$, $\overline{\Upsilon}_{g,g'}$ and $\overline{m}_g$ are the results of fixed point equations with $\boldsymbol{\overline{R}}_g=\boldsymbol{B}_{g}^H\boldsymbol{R}_g\boldsymbol{B}_{g}$: 
\begin{equation}
\overline{m}_g=\frac{1}{b_g}tr(\boldsymbol{\overline{R}}_g\boldsymbol{T}_g)
\label{first}
\end{equation}
\begin{equation}
\boldsymbol{T}_g=\big(\frac{S_g}{b_g}\frac{\boldsymbol{\overline{R}}_g}{\overline{m}_g}+\boldsymbol{I}_{b_g} \big)^{-1}
\end{equation}
\begin{equation}
\overline{\Upsilon}_{g,g'}=\frac{S_{g'}}{b_{g'}}\frac{n_{g',g}}{(\overline{m}_{g'})^2}
\end{equation}
\begin{equation}
n_{g',g}=\frac{\frac{1}{b_{g'}}tr(\boldsymbol{\overline{R}}_{g'}\boldsymbol{T}_{g'}\boldsymbol{B}_{g'}^H\boldsymbol{R}_g\boldsymbol{B}_{g'}\boldsymbol{T}_{g'})}{1-\frac{\frac{S_{g'}}{b_{g'}}tr(\boldsymbol{\overline{R}}_{g'}\boldsymbol{T}_{g'}\boldsymbol{\overline{R}}_{g'}\boldsymbol{T}_{g'})}{b_{g'}(\overline{m}_{g'})^2}}
\label{last}
\end{equation}
As one can see from the equations (\ref{first})-(\ref{last}), the effect of small-scale fading is averaged out and the equations depend only on the channel Second Order Statistics (SOS). This is an interesting and convenient aspect of the equations since the channel SOS change at a much smaller rate than the channel coherence time as previously pointed out in Section \Rmnum
{2}.

An important aspect to conclude from the expressions in (\ref{first})-(\ref{last}) is the fact that the $SINR_{g_k}$ expression depends on the outerprecoder of the group to which user $g_k$ belong to. This entails a combinatorial aspect that complicate scheduling in JSDM in comparison to the typical wireless settings. In fact, by considering the first approach to the outerprecoder, one can clearly see that $B_g$ of each group depends on the activity of other groups and is therefore dependent on the scheduling solution. This will lead to the received power by user $g_k$, denoted by $\frac{P}{S}\zeta_g^2$, to change depending on the groups that are being scheduled. Even the interference that comes from the other groups $\frac{P}{S}\overline{\zeta}_{g'}^2\overline{\Upsilon}_{g,g'}$ change from one scheduling solution to the other. In other words, suppose we have $3$ groups and our aim is to schedule them in order to maximize a certain objective function. If we schedule groups $1$ and $2$ only, the interference from group $1$ to $2$, denoted as $\frac{P}{S}\overline{\zeta}_{1}^2\overline{\Upsilon}_{2,1}$ and from $2$ to $1$, denoted $\frac{P}{S}\overline{\zeta}_{2}^2\overline{\Upsilon}_{1,2}$, is different than the interference between group $1$ and group $2$ is when all groups $1,2,3$ are scheduled simultaneously. Moreover, the received power of users in group $1$ and group $2$, denoted as $\frac{P}{S}\zeta_1^2$ and $\frac{P}{S}\zeta_2^2$ respectively, in the first scheduling solution is different than the latter. In this context, the $SINR_{g_k}$ is made of entities that depend on the scheduling solutions and can be expressed follows:
\begin{multline}
SINR_{g,k}=\\ \frac{x_{g_k}\frac{P}{S}\zeta_g^2(x_1,\ldots,x_{K_G})}{\sum_{g'\neq g}\sum_{k'=1}^{K_{g'}}\frac{P}{S}x_{g_{k'}'}\overline{\zeta}_{g'}^2(x_1,\ldots,x_{K_G})^2\overline{\Upsilon}_{g,g'}(x_1,\ldots,x_{K_G})+1}
\end{multline}
This combinatorial aspect for the channel gains and interference is unique to JSDM and the proposed schemes in the literature for typical wireless scenarios fail here. In fact, this combinatorial aspect that we have when it comes to interference hugely complicate things \cite{7997313} and this is why researchers have been proposing different scheduling solutions to JSDM as seen in \cite{6778043}\cite{7997313}. One question arises: why the focus on this particular choice of outerprecoder $B_g$? The reason behind its importance is the fact that it was shown that the optimal sum-rate scaling law can be achieved by JSDM (the same scaling law of the system capacity with
full channel state information) \emph{under the use} of the approximate block diagonalization \cite{6778043}. Therefore, our goal becomes to study the optimal scheduling under both of those approaches and provide a scheduling scheme that is able to work for either of the outerprecoders choice. This will be tackled in the following. \color{black}

Now that we have dealt with the expression of the achievable rate of each user inside the groups, the goal becomes to schedule these groups in a way to get the highest 
utility, while preserving fairness and ensuring a certain \emph{quality of link} for users inside each group. As for the \emph{quality of link}, we consider as a 
metric the signal to interference Ratio $SIR_{g_k}$ of each \emph{scheduled} user which is a widely used criterion in power control for wireless cellular networks 
\cite{NET-009}. In fact, this criterion is motivated by the well-known physical interference model, where a packet is considered to be successfully transmitted if the 
SIR at the receiver exceeds a certain threshold.
\begin{equation}
SIR_{g_k}=\frac{\overline{\zeta}_g^2}{\sum_{g'\neq g}x_{g'}\overline{\zeta}_{g'}^2\overline{\Upsilon}_{g,g'}}
\label{SIR}
\end{equation}
As seen from the previous expression, the $SIR_{g_k}$ of user $g_k$ depends \emph{solely} on its group index $g$.
%
Putting it all together, and taking the weighted sum-rate as utility, we can formulate our scheduling problem as the following binary optimization problem:
\begin{equation}
\begin{aligned}
& \underset{\boldsymbol{x}\in \{0,1\}^K}{\text{maximize}}
& & \sum_{g=1}^{G}\sum_{k=1}^{K_g}x_{g_k}w_{g_k}R_{g_k}\\
& \text{subject to}
& & SIR_{g_k}\geqslant \alpha_{g_k} \:\:k = 1, \ldots, K_G\\
\end{aligned}
\label{utility}
\end{equation}
The previous studies in this area have not considered fairness which resulted in a portion of users inside the network suffering from starvation (See, e.g.,\cite{7997313}). 
Introducing this weight $w_{g_k}$ allows us to incorporate fairness in our scheduling scheme. An example of a weighting factor choice is the max-weight policy \cite{182479} where the weighting factor $w_{g_k}$ is chosen as the queue length $Q_{g_k}$. As the queue length of user $g_k$ grows larger as a result of not being scheduled, user $g_k$ will have a higher chance of being scheduled on the next time slot. Another example would be to choose the weighting factor $w_{g_k}$ as the inverse of the average achieved rate by user $g_k$ in the previous time slots.
\color{black}

\begin{theorem}
The problem in (\ref{utility}) is NP-hard.
\end{theorem}
\begin{IEEEproof}
See Appendix B.
\end{IEEEproof}
\begin{remark}
The formulation of our problem (\ref{utility}) is not bound to a particular clustering process and therefore the complexity results hold for any users partitioning techniques used.
\end{remark}
With the establishment of the NP-hardness of our problem, one should abandon efforts to find
globally optimal solution and resort to sub-optimal schemes that perform well. 
\subsection{Proposed Scheme}
An interesting approach to propose a well performing sub-optimal scheme is to apply polynomial-time transformations on our problem with the aim of turning it to a well-known problem\footnote{It is worth mentioning that due to the combinatorial aspect previously pointed out to, the standard relaxation of the scheduling, that consists of relaxing the boolean variable constraint to $0\leq x_{g_k}\leq 1$ and recoursing to standard Lagrangian duality theory, fails}. Afterwards, the rich available literature of the resulting problem can be used to adopt approximation algorithms with proven performance guarantees. We therefore proceed to modeling our scheduling problem by using graph theory. Next, we pull together various graph transformations and well-known mathematically established approximation algorithm from the graph theory literature, more specifically from the vertex coloring literature, to construct our scheme. The proposed scheme is made of 3 steps: "Elimination, Grouping, Verification". The first step deals with the $SIR$ constraint. The second step finds the appropriate combination of groups to be scheduled. The third step refines the results of step 2. \\
\textbf{Modeling:} Using the $SIR_{g_k}$ expression in (\ref{SIR}), we can construct a weighted \emph{directed} graph $G_L=(V,E)$ where V is the set of users and in which the weight of the edge $e(g'_{k'},g_k)$ corresponds to what we will call the \emph{normalized interference} from user $k'$ of group $g'$ to user $k$ of group $g$:
\begin{equation}
e(g'_{k'},g_k)=\frac{\overline{\zeta}_{g'}^2\overline{\Upsilon}_{g,g'}}{\overline{\zeta}_{g}^2}
\label{edges}
\end{equation}
An example of 4 groups scenario, each having $1$ user, will be presented in successive figures to demonstrate the mechanisms of the scheduling scheme. We can now proceed with our proposed scheme:
\begin{figure}[!ht]
\centering
\includegraphics[width=.45\linewidth]{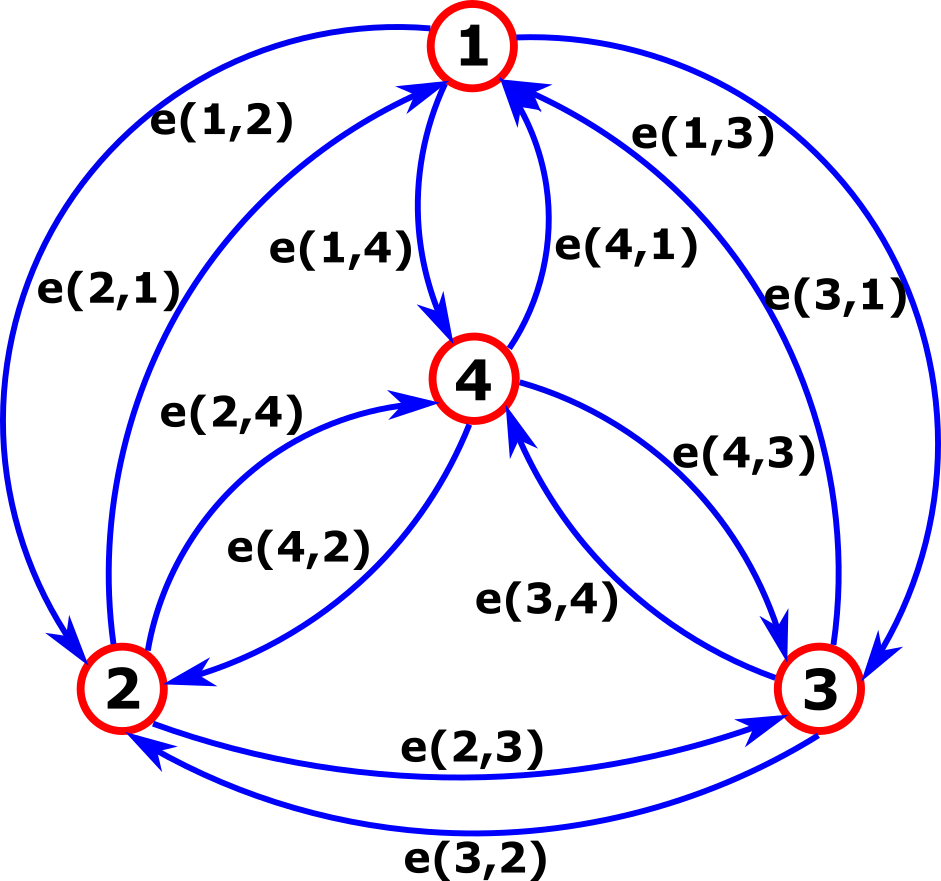}
\setlength{\belowcaptionskip}{-10pt}
\caption{Weighted Directed Graph $G_L$}
\label{feasible}
\end{figure}\\
\textbf{\emph{1) Elimination:}} We distinguish two versions of this step, depending on the outer precoder previously selected:\\
\emph{Approach 1}: 
We can picture each vertex in the graph as a \emph{sink} of interference that undergoes successive \emph{iterations}. In the first iteration, all users are considered 
to be active and the elimination process starts. The outer precoder $\boldsymbol{B}_g$ of each group is calculated as detailed in Section \Rmnum{4}-A. The fixed points 
equations(\ref{first})-(\ref{last}) are then solved and $e_1(g'_{k'},g_k)\; \forall (g'_{k'},g_k)\in V^2$ are calculated based on (\ref{edges}) where the iteration number can be 
visualized in the sub-index "$1$". For each vertex $g_k \in V$, we test the $SIR_{g_k}$ condition\footnote{The model taken into consideration assumes perfect CSI feedback by the users. However, the scheduling scheme presented is not restricted to it. In fact, noisy CSI feedback can be considered by simply adding a residual interference term to the denominator in the SINR expression (\ref{SINR}) (please refer to Appendix A \cite{6542746}). This residual term can be easily taken into account during the elimination phase when testing the $SIR$ condition.} of (\ref{utility}). If it is violated, the edge $e_1(g'_{k'},g_k)$ with the 
highest weight is eliminated. This is equivalent to saying that the group in which users are causing interference to $g_k$ the most is chosen to be eliminated. It is worth mentioning that the interference between users of different groups depend solely on their group index. In other words, suppose we are visiting a vertex $g_k$ belonging to group $g$. If group $g'$ had $2$ users in it and is chosen to be eliminated, we eliminate a random user of those $2$. However, the next time we visit another vertex belonging to the same group $g$, we eliminate the same user of group $g'$ that was previously eliminated (it is straightforward that we will choose users of group $g'$ to be eliminated as all users of groups $g$ are subject to the same interference levels, i.e., users of group $g'$ will be the most interfering to all users from group $g$). This is done for consistency purposes between the visits of users belonging to the same group $g$.

\begin{figure}[!ht]
\centering
\includegraphics[width=.9\linewidth]{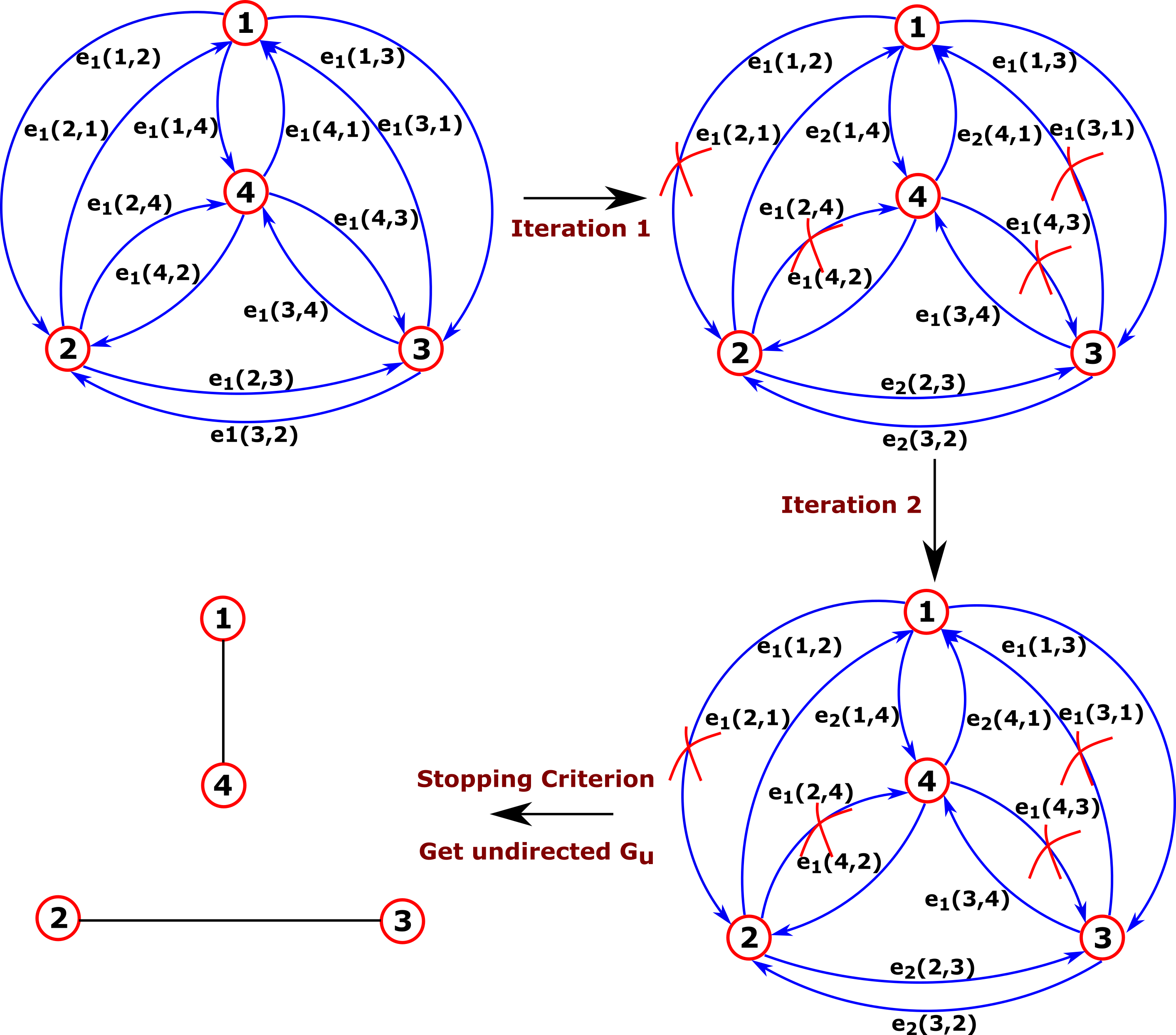}
\setlength{\belowcaptionskip}{-10pt}
\caption{Elimination Process}
\label{eliminationpro}
\end{figure}

\begin{figure}[!ht]
\centering
\includegraphics[width=.9\linewidth]{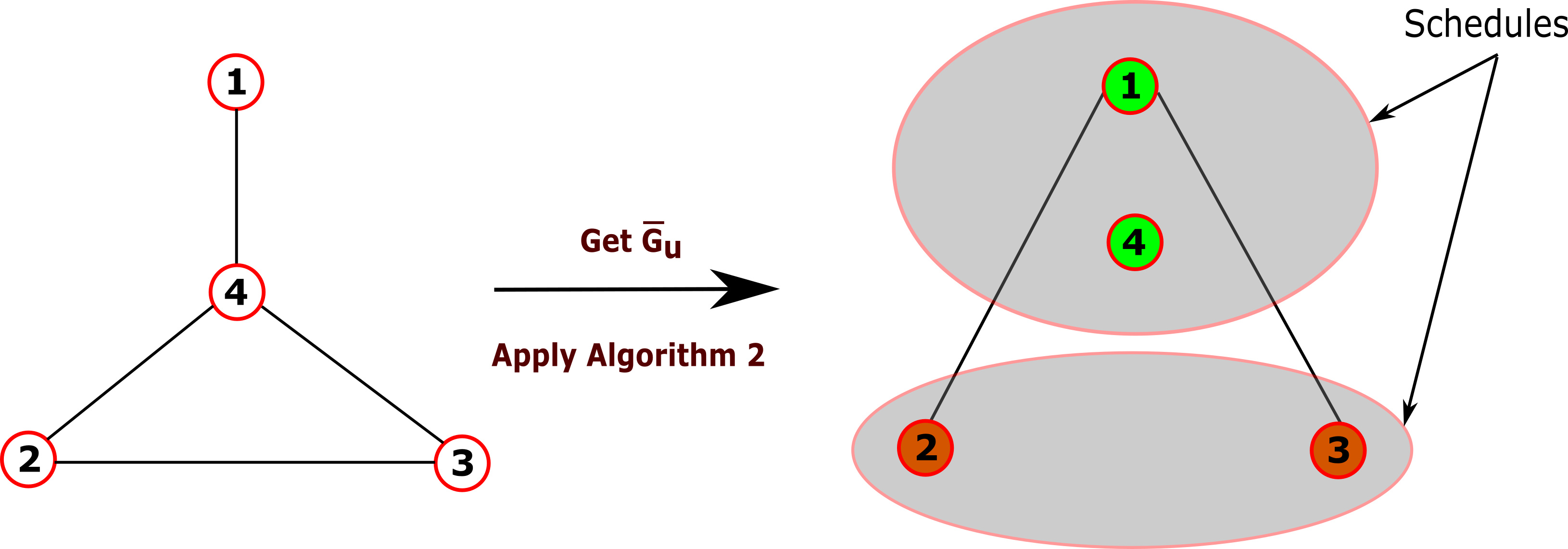}
\setlength{\belowcaptionskip}{-10pt}
\caption{Outliers}
\label{verification}
\end{figure}

At the next iteration, we 
have a new graph due to the edges removal from the previous iteration. This is due to the fact that when adopting Approach 1 for the outer precoder, one can clearly see 
that $\boldsymbol{B}_g$ depends on the eigenspace of all active groups. Hence, when a certain group $g'$ (i.e. all users of group $g'$ have been eliminated) is eliminated from the perspective of users of group $g$, a new 
outer precoder has to be recalculated. Therefore, in this iteration, the outer precoder $\boldsymbol{B}_g$ of each vertex $g_k \in V$ is calculated based on the eigenspace 
of neighboring\footnote{A vertex $g_k$ has group $g'$ as a neighbor at iteration $t+1$ if there exist at least one user $k'$ of $g'$ such that both directed edges $e_t(g_k,g'_{k'})$ and $e_t(g'_{k'},g_k)$ at iteration $t$ were not eliminated} 
groups only. The same procedures take place: the fixed points equations are solved again (\ref{first})-(\ref{last}) and the weight of the edges of \emph{neighboring} 
vertices only are recalculated using (\ref{edges}). The process continues until we reach an iteration that results in no new deleted edges. In this resulting graph, the scheduling of \emph{neighboring} vertices will not violate the corresponding $SIR$ condition of each of the users. An example of the above procedure is presented in Fig. \ref{eliminationpro}, where the first iteration resulted in four deleted edges. The edges of neighboring 
vertices are then updated for the second iteration. The second iteration did not result in any deleted edges and the algorithm finishes. Once it finishes, we turn our directed graph into an \emph{undirected} one $G_u=(V,E_u)$ by simultaneous agreements from both sides i.e. if $e(g_k,g'_{k'})$ and $e(g'_{k'},g_k)$ are both not eliminated in $G_L$ then an 
edge $e_u(g_k,g'_{k'})=1$ exist in $G_u$ and $e_u(g_k,g'_{k'})=0$ otherwise. In our new \emph{undirected} graph $G_u$, an edge exists between two vertices if scheduling them 
together will not violate their respective $SIR$ conditions. \\
\color{black}
\emph{Approach 2}: The elimination step is hugely simplified when employing the second outer precoder approach. Due to the fact that the outerprecoder of each group depends only on its covariance eigenspace, the fixed points equations (\ref{first})-(\ref{last}) are solved only \emph{once} and $e(g'_{k'},g_k)\; \forall (g'_{k'},g_k)\in V^2$ are therefore calculated based on (\ref{edges}). The same procedure stated previously takes place, the only difference is that no recalculation of the edge weights is needed which hugely simplifies the elimination step. Similarly, we will end up with an \emph{undirected} graph $G_u$ where an edge exist between two vertices if scheduling them together will not violate their respective $SIR$ conditions.
\\
\emph{\textbf{2) Grouping:}} After finishing with the elimination step, we can now tackle another aspect of our problem: \emph{Which users of those that are allowed to transmit simultaneously should we 
schedule in order to maximize our utility?} We recall that after proceeding with the elimination step, our $SIR$ constraint can be replaced by making sure that two simultaneously scheduled groups should have an edge between them in $G_u$. Therefore, our problem in (\ref{utility}) is turned into:
\begin{equation}
\begin{aligned}
& \underset{\boldsymbol{x}\in \{0,1\}^K}{\text{maximize}}
& & \sum_{g=1}^{G}\sum_{k=1}^{K_g}x_{g_k}w_{g_k}R_{g_k}\\
& \text{subject to}
& & x_{g_k}+x_{g'_{k'}}\leq 1 \:\:\forall (g_k,g'_{k'})\not\in E_u\\
\end{aligned}
\label{newutility}
\end{equation}
\color{black}
The way we approach this problem is to recall that for a well chosen $\alpha_{g_k} \:\: \forall g{g_k}$, an edge exist between two vertices in $G_u=(V,E_u)$ only if they barely 
interfere and hence scheduling them together would normally increase their sum utility. By taking that into account, our aim becomes to find combinations of users 
that are adjacent one to the other in $G_u$ while covering the whole vertex set $V$. We emphasize the covering aspect of the process to give each group its fair chance 
to access the network. For this purpose, we define a \emph{clique} in an undirected graph as a subset of vertices such that all vertices in the clique are adjacent.
We seek to find the \emph{smallest number} of cliques that cover $V$, where we emphasize "smallest" to ensure that each clique have the largest number of users 
possible inside. Essentially, we are trying to solve the \emph{minimal clique vertex cover} problem. The \emph{minimal clique vertex cover} problem is known to be 
equivalent to the vertex coloring, a well known \emph{NP-complete} problem, on the complement graph $\bar{G}_u$. Knowing that vertex coloring seeks to partition the 
set of vertices into the smallest number of independent sets, one can see the connection between these two problems since a subset of vertices is a clique in $G_u$ if 
and only if it is an independent set in $\bar{G}_u$ (Details on this matter can be found in Appendix A-B). In fact, a graph has a vertex clique cover of size k
\emph{iff} its complement graph can be colored with k colors such that adjacent vertices have different colors. The graph coloring problem is considered one of the most important and the most studied problems
in combinatorial optimization. Due to its importance, the literature is rich with numerous developed polynomial-time algorithms that find approximate solutions for the 
problem with provable guarantees ( see e.g. \cite{Halldorsson2015}). We will therefore use a simple yet effective maximal independent set based vertex coloring 
algorithm that achieves a $O(\frac{n}{log(n)})$-approximation ratio \cite{Halldorsson2015} presented in Algorithm 2 and apply it on $\bar{G}_u$. After applying 
Algorithm 2, each user $g_k$ will be assigned a color $Col(g)$. Groups that are assigned the same color represent a subset of groups that are allowed to transmit 
simultaneously.
We can therefore replace the edges constraint in (\ref{newutility}) by ensuring that the color assignments are respected:
\begin{equation}
\begin{aligned}
& \underset{\boldsymbol{x}\in \{0,1\}^K}{\text{maximize}}
& & \sum_{g=1}^{G}\sum_{k=1}^{K_g}x_{g_k}w_{g_k}R_{g_k}\\
& \text{subject to}
& & x_{g_k}+x_{g'_{k'}}\leq 1 \:\: if \:\: Col(g_k)\neq Col(g'_{k'})\\
\end{aligned}
\label{newnewutility1}
\end{equation}
\color{black}
The problem in (\ref{newnewutility1}) is indeed simple to solve. One can simply form what we will call "\emph{Schedules}", each made of users that belong to the same 
color. These formed schedules are refined in the following step named "Verification".
\begin{figure}[!ht]
\centering
\includegraphics[width=.9\linewidth]{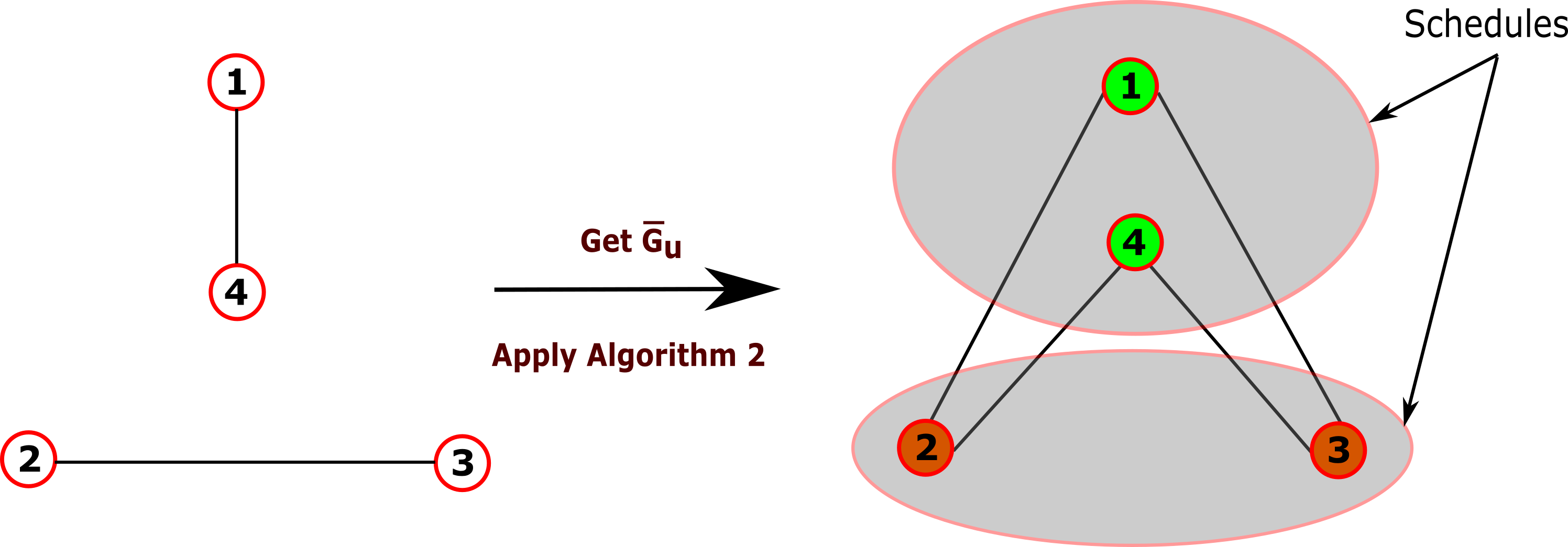}
\setlength{\belowcaptionskip}{-10pt}
\caption{Coloring Process}
\label{color}
\end{figure}\\
\begin{algorithm}
\caption{Coloring Algorithm}\label{euclid}
\begin{algorithmic}[1]
\State \textbf{Initialization:} Let $S=\emptyset$ and index $i=1$
\While {$S\neq V$}
\State Let $C_i=\emptyset$ the set of users assigned the color $i$
\State Let $R=V  \setminus  S $ the set of remaining vertices
\While {$R\neq \emptyset$}
\State Pick a vertex $w\in R$ with lowest degree randomly and let $C_i=C_i \cup \{w\}$
\State $R=R \setminus \{w\}\cup Neighbor(w)$
\EndWhile
\State $S=S \cup C_i$ 
\State Output the color $C_i$ and $i \gets i + 1$
\EndWhile
\end{algorithmic}
\end{algorithm}\\
\textbf{\emph{3) Verification:}} The goal of this step is to refine the results of the previous "Grouping" step. For this purpose, we define \emph{outliers} as users that belong to a certain schedule (i.e. are assigned a specific color) but can be included in others schedules. This means an outlier user can transmit in parallel to groups belonging to a different color and therefore should not be restricted to a single color. An example of an outlier is an isolated group that does not cause any interference on any of the other users. Therefore, it is straightforward that this user should be always transmitting and not restricted to a certain color and hence it should be assigned multiple colors. One can see, for instance in Fig. \ref{verification}, that user "4" belonging to the green schedule have an edge in $G_u$ with each users of the brown schedule. Therefore, user "4" is allowed to transmit simultaneously with groups of the brown color without violating their $SIR$ condition. In other words, user "4" is assigned both colors: green and brown. Hence, each vertex $g_k$ of the graph $G_u$ has to be revisited. If user $g_k$ has an edge with all groups $g'_{k'}$ sharing the same color then user $g_k$ is assigned this additional color. This is done for all vertices and colors and the final schedules are therefore formed.
At the start of each coherence time $T_c$, the schedule that leads to the largest utility is selected. The BS therefore transmits pilots symbols through 
the outerprecoder for the users belonging to this schedule and a measurement of the effective channel is made and fedback to the BS to start the transmission stage.
\section{Numerical Results}
We consider a base station with a $120^\circ$ sector centered around the x-axis consisting of a ULA with $N_t=128$ antennas and serving $K=80$ users arbitrarily  distributed in the sector. As for the correlation entries, we adopt the one-ring model \cite{6542746}. Consider a user terminal (UT) at an azimuth angle $\theta$ and 
angular spread $\Delta$. The correlation entry is then calculated for $1\leqslant m,p\leqslant N_t$ using the following formula:
\begin{equation}
[\boldsymbol{R}]_{m,p}=\frac{1}{2\Delta}\int_{\theta-\Delta}^{\theta+\Delta}e^{j\boldsymbol{k}^T(\alpha)(\boldsymbol{u}_m-\boldsymbol{u}_p)}d\alpha
\end{equation}
where $\boldsymbol{k}(\alpha)=\frac{-2\pi}{\lambda}(cos(\alpha),sin(\alpha))^T$ denotes the wave vector for a planar wave with angle of arrival $\alpha$, $\lambda$ is the wavelength and $\boldsymbol{u}_m,\boldsymbol{u}_p\in \mathbb{R}^2$ are the position vectors of the BS antennas in the 2D-coordinate system. For our scenario, we consider that all users have the same angular spread of $\Delta=5^\circ$. For the upcoming subsections, we suppose that we set the clustering threshold as $DOL_{th}=0.9$. A study on the effect of this threshold is presented in subsection C.
\vspace{-12pt}
\subsection{Scheduling Schemes Comparison}

\begin{figure}[!ht]
\centering
\includegraphics[width=.9\linewidth]{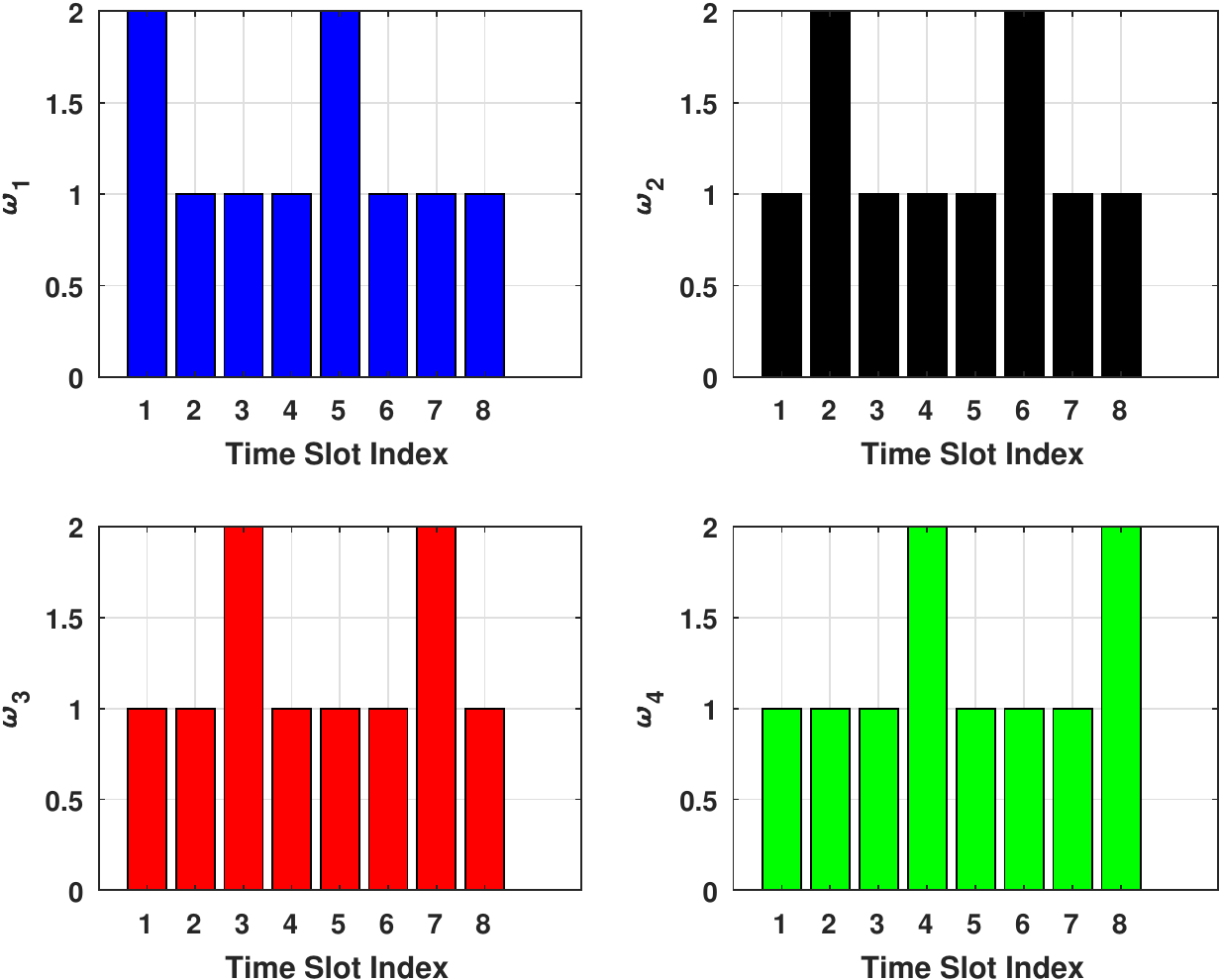}
\setlength{\belowcaptionskip}{-10pt}
\caption{Evolution of the weighting factors}
\label{round}
\end{figure}

\begin{figure}[!ht]
\centering
\includegraphics[width=.85\linewidth]{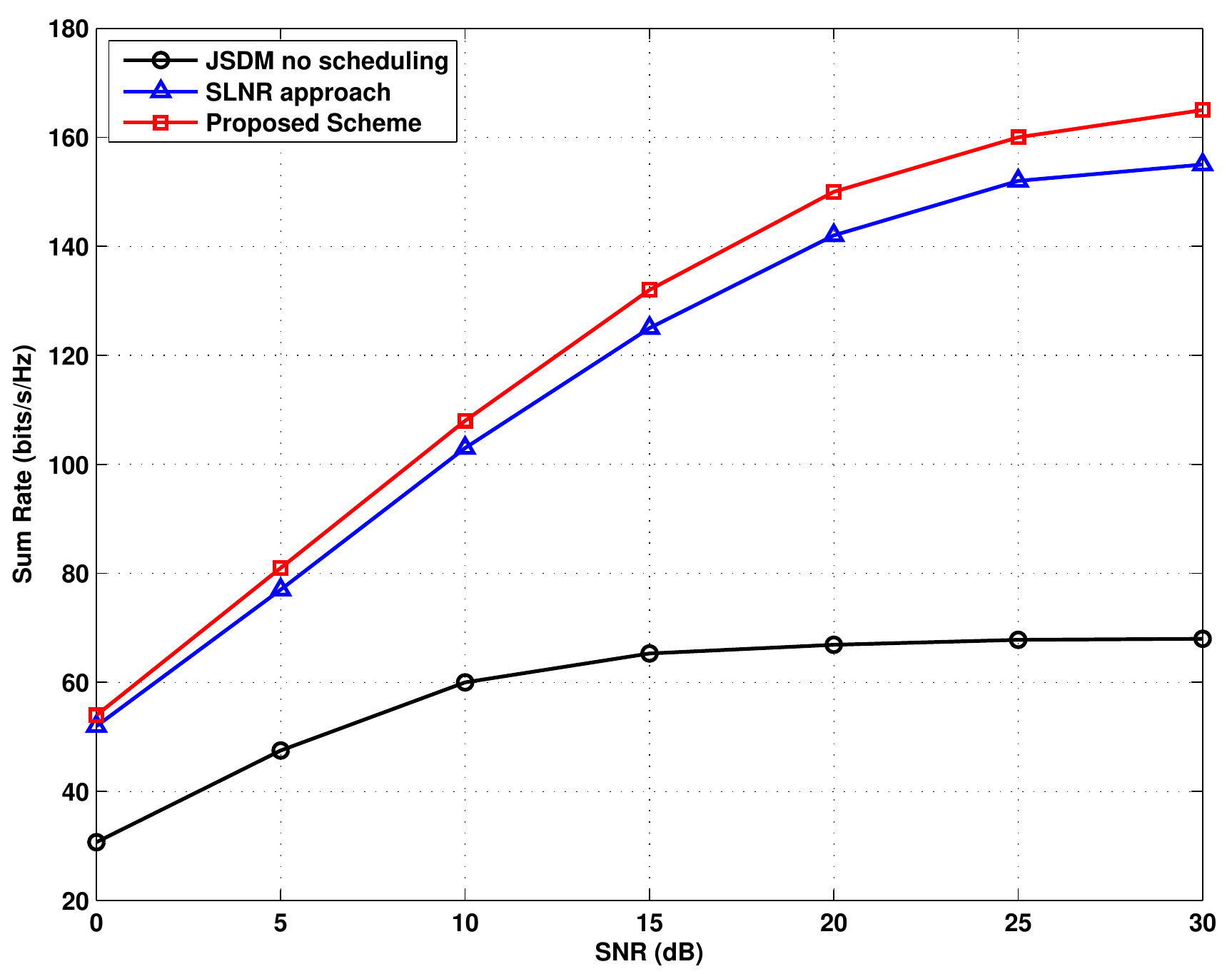}
\setlength{\belowcaptionskip}{-10pt}
\caption{Comparison of sum spectral efficiency vs. SNR}
\label{rate}
\end{figure}

The aim of this simulations section is to compare our proposed scheduling scheme to the recently proposed SLNR based scheduling scheme \cite{7997313}. The SLNR based 
scheduling scheme is taken as a benchmark due to the fact that it was shown to outperform all proposed scheduling methods in the JSDM literature in terms of sum-rate \cite
{7997313}. The importance of adopting a scheduling policy is also highlighted by simulating JSDM without any scheduling just as in \cite{6542746}. 
Due to the fact that both the SLNR based scheduling and our scheduling scheme require a certain threshold tolerance to be set (the SLNR and SIR tolerance 
respectively), we iterate over a wide range of thresholds and choose the one that led to the highest sum-rate as a representative of each method for a fair 
comparison. To compare our scheduling scheme to the SLNR approach counterpart, we create our \emph{schedules} as depicted in Section IV-C. Suppose that we end up having $L$ schedules indexed as $\{1,\ldots,L\}$, we assign the same weighting factor $w_{g_k}$ to the users belonging to the same schedule and update them at each transmission time slot. More precisely, we choose $w_{g_k}$ as follows:
\begin{equation}
  w_{g_k} = \left \{
  \begin{aligned}
    &2 && \text{if}\ n=s_1\:(\text{mod}\:\text{L}),\ldots,n=s_M \:(\text{mod}\:\text{L}) \\
    &1 && \text{otherwise}
  \end{aligned} \right.
\end{equation} 
where $n$ is the transmission time slot, $s_m$ refers to the $m^{th}$ schedule to which user $g_k$ belong to and mod refers to the modulo function. This is a simple weighting factor example as our proposed method is much more general and any desired weighting factors strategy can be adopted. To illustrate this choice of weighting factors, suppose we have $4$ schedules, the evolution of the weighting factors of users belonging to each of the $4$ schedules is detailed in Fig. \ref{round}. The average sum-rate of all users across the \emph{whole transmission time} is taken as a representative of the method. It is worth mentioning that our proposed scheme can achieve an even higher sum-rate by simply constantly (over all time slots) choosing the schedule leading to the highest sum-rate.
\\
As for the outerprecoder, we adopt the first approach and set $r_g^{*}=K_g$ as adopted in \cite{7997313}. Fig. \ref{rate} highlights the fact that JSDM performs poorly without adoping an appropriate scheduling policy due to limitations in terms of interference.
Also, we can see how our proposed scheme was able to outperform the SLNR based method over the whole $SNR=log_{10}(P)$ range. 
The reason behind this is that we work on the interference itself to improve the sum-rate. On the other hand, improving the average SLNR of a system as in \cite
{7997313} does not necessarily translate into a higher sum-rate.

\begin{figure}[!ht]
\centering
\includegraphics[width=.9\linewidth]{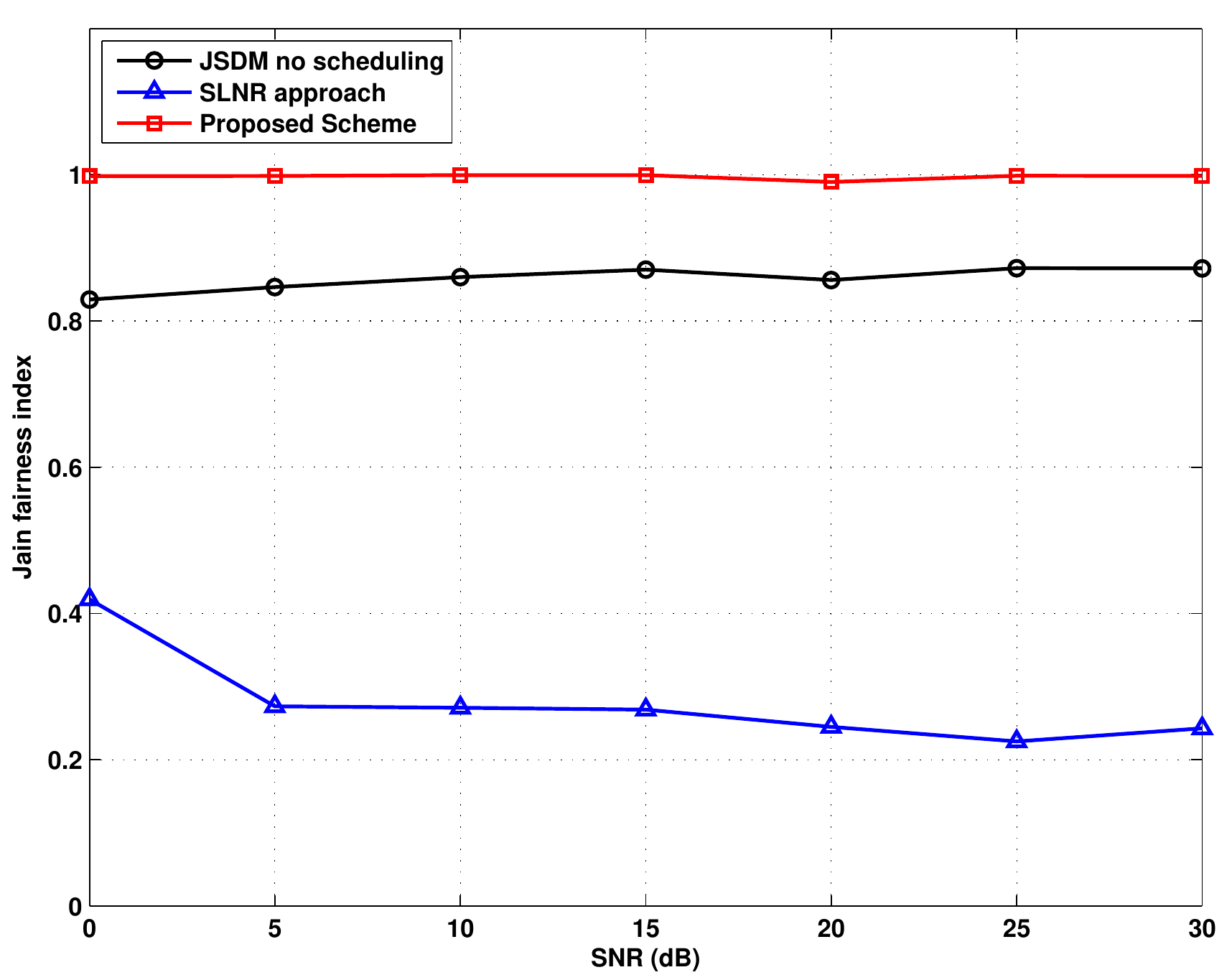}
\setlength{\belowcaptionskip}{-10pt}
  \caption{Comparison of Jain's fairness index vs. SNR}
  \label{fairness}
\end{figure}

With fairness between users being of paramount importance in any scheduling scheme, we consider the well-known Jain's fairness index \cite{6517050} as a metric to 
compare the methods in terms of throughput fairness. It is defined as follows: 
\begin{equation}
1/K\leq F(R_1,R_2,\ldots,R_K)=\frac{(\sum_{k=1}^{K}R_k)^2}{K\sum_{k=1}^{K}(R_k)^2}\leq1
\end{equation}
with $R_k$ being the average rate obtained over the whole time slots as previously explained. The values of this index range from $1/K$ to $1$. \color{black} The lowerbound is achieved when a single user acquires the channel while the others end up starving. As for the 
upperbound, it is achieved when resources are shared equally between users. One can see in Fig. \ref{fairness} how the SLNR based scheduling scored the worst fairness 
index due to the fact that after successive elimination of groups with low SLNR, the users inside these groups end up starving. One can also see how the throughput 
fairness of JSDM with no scheduling is high but not perfect since users suffer different interference conditions and therefore asymmetric throughput.
Our method scored almost perfect throughput fairness due to several reasons: the first being that by construction, the $SIR$ of each group was chosen to be lower bounded by 
the same well chosen tolerance and the second being that symmetrical Round-Robin was adopted between schedules and equal power allocation to all streams was employed. 
\emph{Overall, our proposed scheme was able to outperform the SLNR based method in sum-rate while providing a huge gain in terms of throughput fairness}.

\subsection{Outer precoders Comparison}
\begin{figure}[!ht]
\centering
 \includegraphics[width=.85\linewidth]{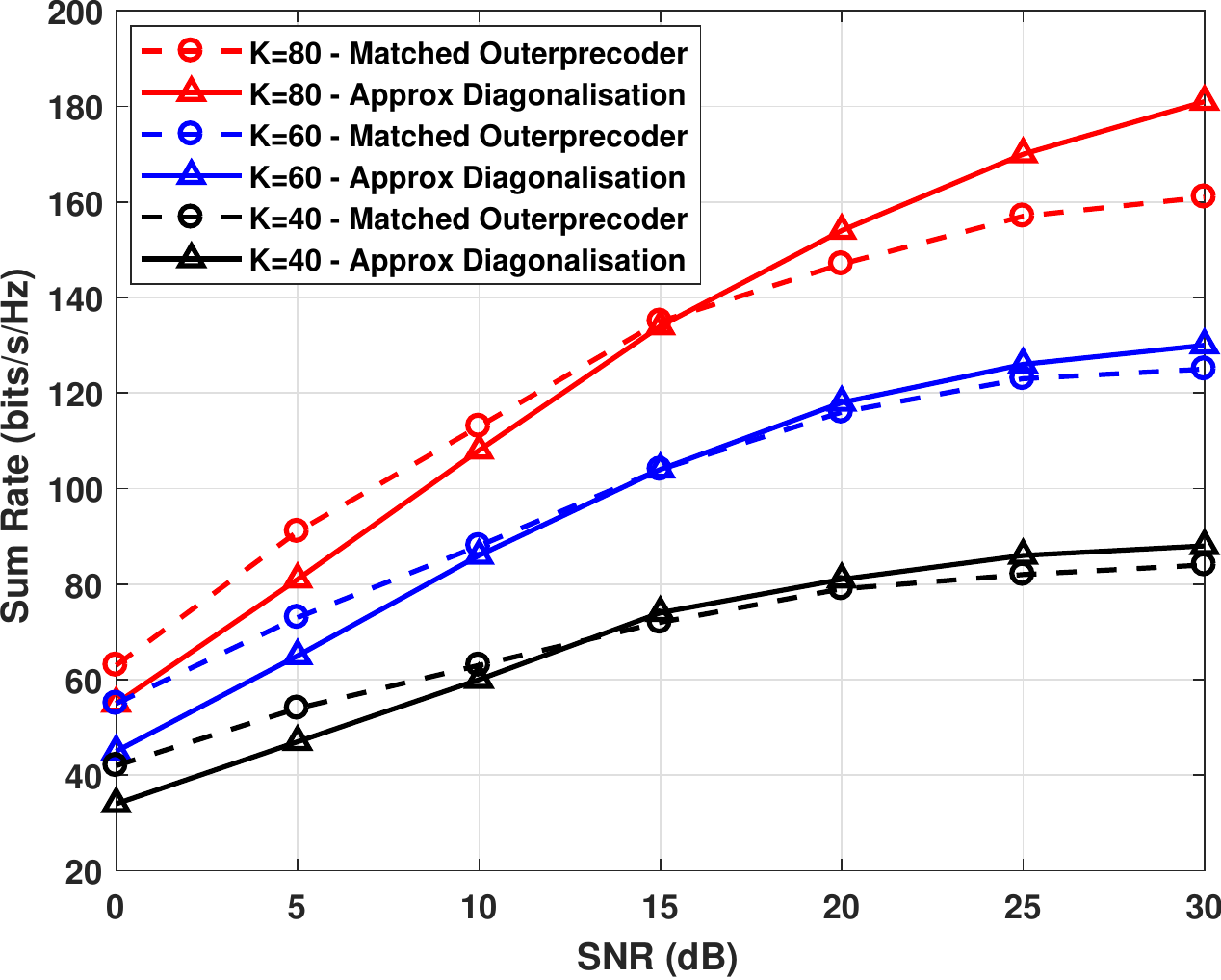}
  \caption{Sum-rate comparison for both outerprecoders vs. SNR}
  \setlength{\belowcaptionskip}{-10pt}
 \label{precoders}
\end{figure}

The goal of these simulations is to compare the two outer precoder approaches discussed in Section \Rmnum{4}. We employ for this purpose our proposed scheduling scheme to compare both approaches. As in our previous scenario, we iterate over a wide range of $SIR$ tolerance and choose the tolerance that led to the highest sum-rate as a representative. By looking at Fig. \ref{precoders}, one can clearly see how the matched outerprecoder outperforms the Approx. Diagonalization, for all the considered total number of users cases, in the $0-15$ dB SNR range before saturating in the high SNR regime. In fact, in realistic scenarios where users overlap in the angular domain, Approx. Diagonalization leads to small channel gains that heavily degrade performance. However, this low channel gain can be easily overcome in the high SNR regime. In this regime, the Approx. Diagonalization outerprecoder is able to outperform the matched outerprecoder counterpart since inter-group interference is canceled at a low performance penalty. One can therefore argue that, in certain scenarios, it is better to deal with inter-group interference \emph{solely} at the MAC layer rather than the combination of PHY/MAC Layer in JSDM for the realistic SNR regime. These results are of paramount importance since the use of the matched outerprecoder hugely simplifies the scheduling procedure due to elimination of the combinatorial aspect found in the first outerprecoder design.  \color{black}
\subsection{Optimal Clustering Threshold}
As it has been previously stated, the performance of JSDM is highly influenced by the clustering solutions obtained and therefore a discussion on the optimal clustering threshold is pivotal to the work. The difficulty in determining the optimal clustering threshold $DOL_{th}$ comes from the fact that it depends on a large number of factors. Some non-exclusive examples include the number of antennas, the number of users and their respective covariance space. Even the power budget taken in consideration can have a huge impact on the optimal threshold. This is a result of a fundamental trade-off that is highlighted when attempting to choose the clustering threshold $DOL_{th}$. In fact, by choosing $DOL_{th}$ too high, we end up with a perfect representative of each group's covariance. This will make the outer precoding techniques perform extremely well. However, by doing so, groups become of smaller size and the gain from having more users in each group and suppressing intra-group interference by inner precoding techniques vanishes. This will create a \emph{burden} on the inter-group interference techniques (both PHY and MAC techniques) to overcome this high inter-group interference. On the other hand, by choosing $DOL_{th}$ too small, groups grow larger but any outerprecoding techniques, both Approaches 1 and 2, will start failing due to the fact that each group's equivalent covariance is not well represented. This trade-off is of paramount importance for the overall performance of JSDM and one has to choose values between the two extremes. 

\begin{figure}[!ht]
\centering
\includegraphics[width=0.85\linewidth]{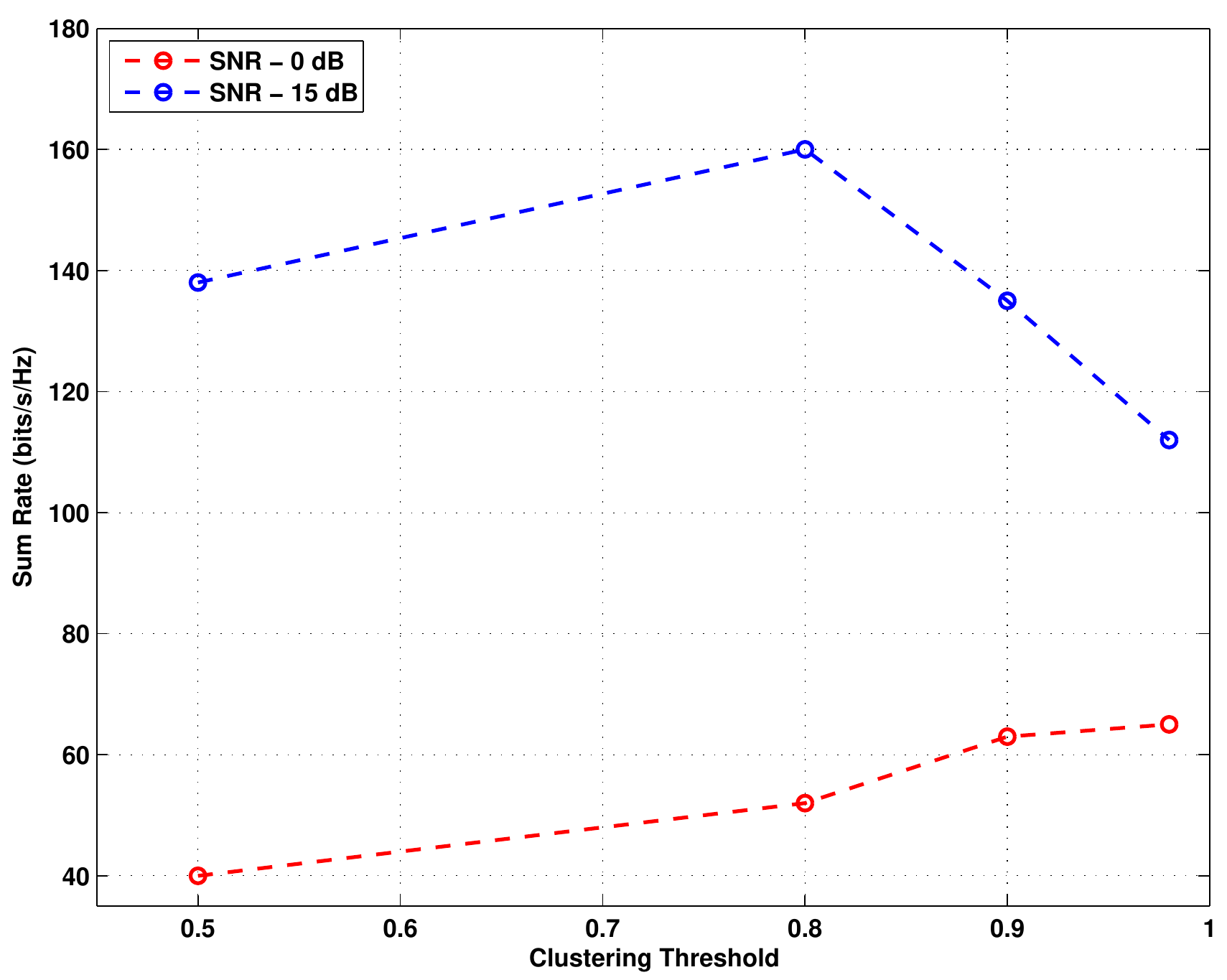}
\setlength{\belowcaptionskip}{-10pt}
\caption{Sum-rate comparison for different $DOL_{th}$}
\label{threshold}
\end{figure}

To visualize this trade-off, we consider a scenario where we use the matched outerprecoder and two SNR conditions. The results are shown in Fig. \ref{threshold}. In the low SNR regime, the scheme is not saturated by inter-group interference, and therefore extracting the maximum possible signal is what matters. Hence, one can clearly see how the performance grows with the clustering threshold due to the fact that groups covariance matrices are well represented and being properly matched when the threshold goes higher. In the high SNR regime (the scenario where saturation by inter-group interference took place in the previous section), the performance is bad for low threshold values since outerprecoding techniques are failing. The performance increases with the threshold before starting to decrease the higher the threshold values goes due to saturation in terms of inter-group interference. One has to therefore choose a threshold between these two extremes to have an overall good performance.
\\
To address this difficulty, we capitalize on four points:
\begin{enumerate}
\item the proposed similarity measure's search space is small (recall that the similarity measure is upperbounded by $1$ and lowerbounded by $0$)
\item the clustering scheme runs in polynomial time
\item the rate of each user can be easily approximated using the expression in (\ref{SINR})
\item the second order statistics of the channel vary at a much slower rate than the channel's coherence time
\end{enumerate}
We can therefore initially start with a clustering threshold $DOL_{th}(0)=1$ and proceed to apply the clustering algorithm and scheduling scheme. The objective function is then evaluated and the clustering threshold $DOL_{th}$ is decremented by a small step $\delta$ as depicted in the following:
\begin{equation}
DOL_{th}(i+1)=DOL_{th}(i)-\delta
\end{equation}
where $i$ is the iteration number. If the evaluated objective function has decreased, the procedure is stopped and $DOL_{th}(i)$ is taken as the clustering threshold to be used until the second order statistics of the users change. To illustrate the advantage of our proposed clustering measure, we provide a comparison with the chordal distance approach adopted in \cite{6778043}\cite{7997313}. We consider a user at an angle $\theta_0$, having an angular spread of $\Delta=5^{\circ}$ and we calculate both the chordal distance and our proposed measure between its covariance space and the spaces of several different users in the cell. One can clearly see that, for a fixed $\theta_0$, the range of the chordal distance is huge in comparison to our proposed metric which makes it difficult to find a threshold for which the performance is optimal. One can also notice that for different values of $\theta_0$, the chordal distance hugely fluctuates even if $\theta-\theta_0$ is kept constant which is an unappealing property in terms of threshold design. Combining all those observations, we can conclude that it is easier to find optimal thresholds for our proposed measure due to the small search space in comparison to the chordal distance.
\begin{center}
\begin{tabular}{|c|c|c|c|c|}
 \hline
 & \multicolumn{4}{|c|}{Clustering Measures} \\
 \hline
  $\theta_0$ & $\theta-\theta_0$ & $\Delta$ & Chordal Distance & Proposed Measure \\
  \hline
 & $1^{\circ}$  & $5^{\circ}$  & $200$  & $0.9280$\\
 $0^{\circ}$    & $3^{\circ}$    & $5^{\circ}$  & $760$  & $0.7275$\\
 & $30^{\circ}$     & $5^{\circ}$  & $2956$  & $7\times10^{-4}$ \\
 \hline
  & $1^{\circ}$  & $5^{\circ}$  & $220$  & $0.9313$\\
 $30^{\circ}$    & $3^{\circ}$    & $5^{\circ}$  & $871$  & $0.7311$\\
 & $30^{\circ}$     & $5^{\circ}$  & $4126$  & $9\times10^{-4}$ \\
 \hline
\end{tabular}
 \captionof{table}{Clustering Distances}
 \vspace{-4pt}
\end{center}


\color{black}
\section{Conclusion}
In this paper, we tackled the problem of users clustering and scheduling in the promising technique of two-stage beamforming for the downlink of FDD massive MIMO. We introduced a new similarity metric coupled with a clustering method that are characterized by ease of design and good performance. We also presented how in certain scenarios, inter-group interference is better to be dealt with solely on the MAC layer. 
We provided fundamental complexity results for finding the optimal scheduling policy in JSDM by proving it to be NP-hard. Knowing that a polynomial time algorithm to solve optimally our scheduling problem is unfeasible, unless P=NP, we developed an efficient scheduling scheme based on graph theory. The proposed scheme was shown to outperform currently available methods in the literature in both sum-rate and throughput fairness.

\bibliographystyle{IEEEtran}
\bibliography{trialout}
\appendices
\section{Required Background}
\subsection{Clustering Algorithm}
The clustering scheme presented in Section III has been formulated as a binary optimization problem that is proven to be NP-Hard. To address this difficulty, the literature tackles the problem by relaxing the binary condition and solving the linear program (\ref{LP}). To proceed with the mapping of the LP solutions to the  binary problem, the pivoting procedure has been introduced. Pivoting works by treating the fractional solutions of the LP as a probability to put the two vertices in 
different clusters. The algorithm that was proposed in \cite{Chawla:2015:NOL:2746539.2746604} is to apply the following functions on the solution of (\ref{LP}) before 
proceeding to the pivoting phase:
\begin{equation}
f^+(x_{uv}) = \begin{cases}
             0  & \text{if } x < a \\
             (\frac{x_{uv}-a}{b-a})^2  & \text{if } x \in [a,b] \\
             1  & \text{if } x \ge b
       \end{cases} \quad
f^-(x_{uv}) = x_{uv}
\label{funct}
\end{equation}
where $\langle+\rangle$ and $\langle-\rangle$ refer to $(u,v) \in E_c^+$ and $(u,v) \in E_c^-$ respectively. This rounding technique is guaranteed to achieve an expected (2.06-$\epsilon$)-approximation for $a=0.19$, $b=0.5095$ and a constant $\epsilon$ such as $0<\epsilon<0.01$. A derandomized version of the algorithm was also proposed in \cite{Chawla:2015:NOL:2746539.2746604}, at the cost of increased complexity, but we omit it for the sake of space and we refer the readers to \cite{Chawla:2015:NOL:2746539.2746604} for a more detailed discussion. 
\subsection{Graph Theory}
The purpose of this subsection is to introduce necessary graph theory clarify the equivalence between the two well-known graph problems: minimal vertex clique cover and graph vertex coloring. We first start by defining a clique of an undirected graph $G=(V,E)$.
\begin{definition}
A clique is a subset of vertices of an undirected graph $G$ such that every two distinct vertices in the clique are adjacent; that is, its induced subgraph is complete. 
\end{definition}
Armed with this definition, we define the minimal vertex clique cover along with the graph coloring problem and clarify the equivalence between the two.
\begin{definition}
Given an undirected graph $G$, a vertex clique cover is a partition of the vertices of the graph into cliques. A minimal vertex clique cover is a vertex clique cover that uses as few cliques as possible.
\end{definition}
\begin{definition}
Given an undirected graph $G$, a proper vertex coloring is an assignment of colors to each of the graph's vertices such that adjacent vertices receive different colors. The graph coloring problem aims at minimizing the number of colors used with the chromatic number $\chi(G)$ being the minimum number of colors that can be used.
\end{definition} 
It is well known in the graph theory literature that these two problems are equivalent. To see this more clearly, we consider the decision version of each of the problem in question. In other words, answering the following question: Given an integer $k$ and an undirected graph $G$, can the graph's vertices be covered by at most $k$ cliques? The answer to this question is TRUE if and only if the answer to the following question is TRUE: Given an integer $k$ and an undirected graph $G$, can the complement graph $\bar{G}$ be colored with at most $k$ colors? This is a natural conclusion from the fact that a subset of vertices is a clique in $G$ if 
and only if it is an independent set in $\bar{G}_u$. 
\section{Proof of NP-hardness}
The standard method to show that a certain optimization problem is NP-hard is to establish the NP-hardness of its corresponding decision problem. The decision version of our problem in (\ref{utility}) is to answer by TRUE or FALSE the following question: is there a scheduling solution such that the overall weighted sum-rate is larger or equal to a certain number, say $\gamma$? One can see clearly that the decision version of our problem is easier than the problem in (\ref{utility}), since the latter further requires finding the global maximal value and maximizer. Therefore, if we establish that the decision version is NP-hard then our problem in (\ref{utility}) is itself NP-hard. 
In complexity theory, to prove that a certain decision problem $A$ is NP-hard, we first have to choose a well-know NP-complete problem $B$. Afterwards, we construct a polynomial time reduction from any instance of this problem $B$ to a particular instance of our problem $A$. Under this reduction, the answer to problem $B$ should be TRUE \emph{if and only if} that particular instance of problem $A$ is itself TRUE. The main difficulty lies in finding that suitable NP-complete problem $B$ along with the appropriate scenario of problem $A$ where the reduction takes place. For our case, the proof is based on a polynomial reduction from the SAT problem, the first proven NP-complete problem and the most widely used problem to prove NP-hardness.

The Boolean satisfiability problem, commonly abbreviated as SAT, is the problem of establishing that a certain Boolean formula can be evaluated as TRUE by assigning the values TRUE or FALSE to the associated Boolean variables. Taking into account that SAT is the basis of our proof, we therefore present the common SAT terminology. Let $\boldsymbol{x}=\{x_1,x_2,\ldots,x_M\}$ be a Boolean vector of size $M$. We define conjunction and disjunction, denoted by $\vee$ and $\wedge$ respectively, as the binary AND and OR operators respectively. We also define $\bar{x}_m$ as the logical complement of $x_m$. The binary variables $(x_m,\bar{x}_m) \:\:\: \forall m$ are referred to as \emph{literals}. A \emph{clause} $d$ is defined as a disjunction of literals (e.g. $d=x_1\vee x_2\vee x_3$). A formula $F$ is in Conjunctive Normal Form (\textbf{CNF}) if it is a conjunction of several clauses (e.g. $F=(x_1\vee x_2\vee x_3)\wedge(\bar{x}_4\vee x_5\vee x_6)$). It is well known by the Boolean algebra laws that every logical formula can be transformed into a CNF. This is the reason why SAT seeks an assignment vector $\boldsymbol{x}$ such that a certain CNF formula F can be evaluated as TRUE.
The SAT instance used in the proof is supposed to be irreducible. A SAT instance is said to be irreducible if the elimination of any literal or clause is non-trivial. In other words, we suppose that the following cases do not take place:
\begin{itemize}
\item Only one of $x_m$ or $\bar{x}_m$ appears in F, then it is trivial to assign the value TRUE and FALSE respectively to the binary variable $x_m$.
\item A clause is trivially satisfied (e.g $d=(x_1\vee \bar{x}_1\vee x_3)$) and therefore can be eliminated from F.
\end{itemize}
With the terminology dealt with, we can now proceed to our proof. 

We consider a SAT problem defined as a conjunction of $D$ clauses over a boolean vector $\boldsymbol{x}$ of size $M$. To establish a mapping between our problem and this SAT instance, we consider two sets of groups, the first being $\mathbb{M}\triangleq\{1,\ldots,M,\bar{1},\ldots,\bar{M}\}$ and the second being $\mathbb{D}\triangleq\{1,\ldots,D\}$ mapped to the $2M$ literals and $D$ clauses of the SAT instance respectively. In other words, if the group mapped to literal $m$ is scheduled then this means $x_m=TRUE$. Also, if the clause group $d$ is scheduled then $d=TRUE$. The proof works for any number of users per group which makes it independent of the clustering technique employed. However, for the sake of simplicity, the number of users in each group is set to $1$ and therefore the word user will be used instead of group in the sequel. We suppose that the weight for each user
is set to be equal to $1$ and the objective function becomes the sum-rate.
We rewrite the $SIR$ condition as follows: 
\begin{equation}
SIR_{i}=\frac{G_{ii}}{\sum_{j\neq i, j\in L }G_{ji}} \quad \quad \forall i \in L
\label{SIR}
\end{equation}
where $L$ is the set of all users. We can now proceed with the construction of our special instance of problem (\ref{utility}). In order for our original problem to comply with the SAT instance requirement, we first suppose that $G_{mm}=G_{\bar{m}\bar{m}}=\rho \:\: \forall m,\bar{m} \in \mathbb{M}$. We also suppose that $G_{dd}=\beta \:\: \forall d\in \mathbb{D}$.  We also consider that:
\begin{equation*}
G(m,m') = \begin{cases}
           \rho  & \text{if} \:\: m'=\bar{m}\:\: \text{or}\: \: m=\bar{m}' \\
           0  & \text{otherwise} 
       \end{cases} \quad
\end{equation*}
We consider the case where users of the clause set $\mathbb{D}$ do not interfere i.e. $G_{ij}=0 \:\: \forall \: i,j\in \mathbb{D}$. As for the interaction between the two sets, we suppose that if user $m$ is a literal of the clause associated to user $d$, then $G_{md}=G_{dm}=0$ otherwise $G_{md}=\frac{1}{M}$ and $G_{dm}=\delta$. We suppose that $D\delta<\rho$, and set the quality of link tolerance of the literals set to be $\alpha_{1}=1+\epsilon_{1} \:\: \forall m,\bar{m} \in \mathbb{M}$ with $\epsilon_{1}$ is a \emph{strictly} positive number that satisfies $\epsilon_{1} \leq \frac{\rho-D\delta}{D\delta}$. This special case of mutual interference and quality of link tolerance ensures that users $m$ and $\bar{m}$ cannot be simultaneously scheduled since scheduling them together will violate the $SIR$ 
requirements of each one of them. In other words, the value TRUE cannot be assigned to their mapped literals simultaneously which coincides with the Boolean variables 
requirements. Also, the condition of $\epsilon_{1}$ ensures that the scheduling of clauses users do not impose any restriction on the scheduling of the literals users. As for the clauses users, we set the quality link tolerance as $\alpha_{2}=\beta+\epsilon_{2} \:\: \forall d\in \mathbb{D}$ where 
$\epsilon_{2}$ is a \emph{strictly} positive number that satisfies $\epsilon_{2} \leq \frac{\beta}{M-1}$.
As we can see from this condition, a user $d$ from the clauses set can be scheduled if and only if at least one of its literal users is scheduled as well. Otherwise, $SIR_d=\frac{\beta}{M(\frac{1}{M})}=\beta<\alpha_2$. Also, one can see that in the case when exactly one literal of clause $d$ is scheduled, $SIR_d=\frac{\beta}{(M-1)(\frac{1}{M})}\geq \alpha_2$ due to the imposed condition on $\epsilon_2$. These assumptions coincide with the disjunction nature of each clause. Therefore, with this interference setting, our problem can be successfully mapped to the SAT instance. The question to answer now is: are the preceding interference settings a plausible scenario of our original problem? It is worth mentioning that the preceding scenario is not bound to a particular power allocation scheme nor to a specific covariance model like the one-ring model used in the simulations.

To prove so, we start first with a discussion on the effect of Large scale fading (path loss). Suppose we have a user $g_1$ with covariance $\boldsymbol{R}_{g_1}$. Due to Large-scale fading, the covariance of this user scales down to $\boldsymbol{R}_{g_1'}=\kappa\boldsymbol{R}_{g_1}$ with $\kappa<1$. One can clearly see from the fixed points equation that due to the linearity of the trace function, the following holds:
\begin{equation}
\overline{\zeta}_{g_1'}^2=\kappa\overline{\zeta}_{g_1}^2 \quad \overline{m}_{g_1'}=\kappa\overline{m}_{g_1} \quad \boldsymbol{T}_{g_1'}=\boldsymbol{T}_{g_1} \quad \overline{\Upsilon}_{g_2,g_1'}=\frac{1}{\kappa}\overline{\Upsilon}_{g_2,g_1}
\end{equation} 
From the previous equation, one can see that the same interference caused on other users $g_2$ stays the same ($\overline{\zeta}_{g_1'}^2\overline{\Upsilon}_{g_2,g_1'}=\overline{\zeta}_{g_1}^2\overline{\Upsilon}_{g_2,g_1}$). However, the interference from other users on itself change: 
\begin{equation}
\overline{\zeta}_{g_2}^2\overline{\Upsilon}_{g_1',g_2}=\kappa\overline{\zeta}_{g_2}^2\overline{\Upsilon}_{g_1,g_2}
\end{equation} 
This can be justified since if two users $1$ and $2$ share common modes but user $2$ is suffering from severe path loss, the interference from the link of user $1$ to user $2$ is heavily reduced. However, the interference from user $2$ on user $1$ remains high. Therefore, the differences in path-loss can be one of justification of asymmetric interference levels between literals/clauses users. We therefore suppose the following:
\begin{enumerate}
\item $\forall m,m'\in\mathbb{M}$ such as $m'\neq\bar{m}$, literals users $m,m'$ have orthogonal covariance matrices
\item All literals users $m$ have shared modes with their corresponding literals users $\bar{m}$
\item Each clause user share only a small number of modes with its non-corresponding literals users while its other modes are orthogonal to its corresponding literals
\item The clauses users do not share any common modes
\end{enumerate}
The first condition ensures that literals user $m$ do not interfere on other literals $\forall m'\neq \bar{m}$. Condition 2 ensures that literals $m$ and $\bar{m}$ highly interfere. The third condition ensures that clauses users and their corresponding literals do not interfere however they interfere with their non-corresponding literals. The fourth condition ensures that clauses do not interfere. An example for $M=3$ and $D=3$ can be presented by taking into account $N_t$ orthogonal vectors $\Omega=\{\boldsymbol{u}_1,\ldots,\boldsymbol{u}_{N_t}\}$ that form a basis of the $N_t$ dimensional space where $N_t=128$ is the number of antennas. Consider the following case for the literal users:
\begin{equation}
\boldsymbol{R}_1=\boldsymbol{U}_1\boldsymbol{\Lambda}_1\boldsymbol{U}_1^{H} \quad \boldsymbol{R}_2=\boldsymbol{U}_2\boldsymbol{\Lambda}_2\boldsymbol{U}_2^{H} \quad \boldsymbol{R}_3=\boldsymbol{U}_3\boldsymbol{\Lambda}_3\boldsymbol{U}_3^{H}
\end{equation}
where $\boldsymbol{U}_i=[\boldsymbol{u}_{20(i-1)+1},\ldots,\boldsymbol{u}_{20(i-1)+20}] \in \mathbb{C}^{N_t\times 20}$ and $\boldsymbol{\Lambda}_i$ is a diagonal matrix with the channel modes gain as diagonal entries. The effect of path loss is incorporated inside $\boldsymbol{\Lambda}_i$. This case can be justified since in massive MIMO settings, the covariance of each user is of rank $r_g\ll N_t$ \cite{6542746} which is taken as $20$ and $16$ for literals and clauses users respectively. Similarly for the complement literals users:
\begin{equation}
\boldsymbol{R}_{\bar{1}}=\boldsymbol{U}_{\bar{1}}\boldsymbol{\Lambda}_{\bar{1}}\boldsymbol{U}_{\bar{1}}^{H} \quad \boldsymbol{R}_{\bar{2}}=\boldsymbol{U}_{\bar{2}}\boldsymbol{\Lambda}_{\bar{2}}\boldsymbol{U}_{\bar{2}}^{H} \quad \boldsymbol{R}_{\bar{3}}=\boldsymbol{U}_{\bar{3}}\boldsymbol{\Lambda}_{\bar{3}}\boldsymbol{U}_{\bar{3}}^{H}
\end{equation}
where $\boldsymbol{U}_{\bar{i}}=[\boldsymbol{U}'_i,\boldsymbol{U}''_i] \in \mathbb{C}^{N_t\times 20}$ with $\boldsymbol{U}'_i$ being a set of $10$ eigenvectors from literal user $i$ while $\boldsymbol{U}''_i$ is a set of $10$ eigenvectors non used from the pool $\Omega$. As for the literals users, suppose we have $d_1=x_1\vee x_2\vee x_3$, then simply we can consider the case where $\boldsymbol{R}_{d_1}=\boldsymbol{U}_{d_1}\boldsymbol{\Lambda}_{d_1}\boldsymbol{U}_{d_1}^{H}$ where $\boldsymbol{U}_{d_1}=[\boldsymbol{U}'_{\bar{1}},\boldsymbol{U}'_{\bar{2}},\boldsymbol{U}'_{\bar{3}},\boldsymbol{U}''_{d_1}]$ where each $\boldsymbol{U}'$ is made of $2$ eigenvectors from the corresponding literals users while $\boldsymbol{U}''$ is made of $10$ eigenvectors from the pool $\Omega$. Obviously, this setting verifies all the cited conditions earlier and therefore approves that the 
interference setting taken into account is indeed a plausible case of our original problem. We therefore proceed with the rest of the proof.

It is important now to discuss the scenario that would lead to the maximal objective function possible. We know that the maximum number of scheduled literals users is $M$ since two complement literals are forbidden to be scheduled at the same time. As for the clauses users, the maximum number is $D$. Suppose that the maximization of the objective function in this setting leads to $k_1$ users from the literals set to be scheduled. The scheduling of these $k_1$ literals would \emph{allow} us to schedule $k_2$ users from the clauses without violating their $SIR$ conditions. Suppose we decide to schedule $k'_2$ out of these $k_2$ allowed clauses users.
\begin{lemma}
If $log_2(1+\frac{\beta}{1+\frac{M-1}{M}})+Mlog_2(1-\frac{\rho\delta}{(1+\rho)(1+\delta)})\geq0$, then the objective function increases with $k'_2$ and therefore we should choose $k'_2=k_2$
\label{lemma1}
\end{lemma}
\begin{IEEEproof}
We consider our objective function for $k_1$ literals users and $k'_2$ clauses users:
\begin{equation}
T(k_1,k'_2)=\sum_{i=1}^{k_1}log_2(1+\frac{\rho}{1+g_i(k'_2)\delta})+\sum_{j=1}^{k'_2}log_2(1+\frac{\beta}{1+\frac{f_j(k_1)}{M}})
\end{equation}
where $g_i(k'_2)$ and $f_j(k_1)$ refer to the number of interfering clauses and literals users of those that are scheduled respectively. To study the effect of scheduling an additional clause user of those we are allowed to schedule, we take the marginal gain as a basis for our analysis. The worst case for the gain takes place when this additional added clause user add interference on all the literals users previously scheduled and this added clause user suffers from the worst possible interference. In other words, $g_i(k'_2+1)=g_i(k'_2)+1 \:\: \forall i$ and $f_{k'_{2}+1}(k_1)=k_1-1$. The lower-bound on the gain becomes:
\begin{equation}
\begin{aligned}
& G=T(k_1,k'_2+1)-T(k_1,k'_2)\geq log_2(1+\frac{\beta}{1+\frac{k_1-1}{M}})+\\
&\sum_{i=1}^{k_1}log_2(\frac{1+\rho+\delta+\delta g_i(k'_2)}{1+\delta g_i(k'_2)+\delta}\frac{1+\delta g_i(k'_2)}{1+\rho+\delta g_i(k'_2)})
\end{aligned}
\end{equation}
The expression inside the second term takes the form of $\frac{(a+\delta)b}{(b+\delta)a}=1+\frac{\delta(b-a)}{ab+a\delta}$ with $a=1+\rho+\delta g_i(k'_2)$ and $b=1+\delta g_i(k'_2)$. Knowing that $g_i(k'_2)\geq0$, we have:
\begin{equation}
G\geq log_2(1+\frac{\beta}{1+\frac{k_1-1}{M}})+k_1log_2(1-\frac{\rho\delta}{(1+\rho)(1+\delta)})
\end{equation}
It is enough that this lower bound to be positive $\forall k_1\leq M$ for the objective function to be increasing with $k'_2$. Instead of checking $M$ conditions, we define the following function: $f(x)=log_2(1+\frac{\beta}{1+\frac{x-1}{M}})+xlog_2(1-\frac{\rho\delta}{(1+\rho)(1+\delta)})$. By deriving with respect to $x$, we have $f'(x)=\frac{1}{ln(2)}\frac{\frac{-\beta/M}{(1+\frac{x-1}{M})^2}}{1+\frac{\beta}{1+\frac{x-1}{M}}}+log_2(1-\frac{\rho\delta}{(1+\rho)(1+\delta)})$.
One can easily verify that $f'(x)$ is negative $\forall x\geq1$. This tells us that $f(x)$ is actually decreasing with respect to $x$ and therefore if $f(M)\geq0$ for a certain $M\in \mathbb{N}^{*}$, then $f(k_1)\geq0\:\: \forall 
k_1\leq M$. Therefore, it is enough for that condition to be verified for $k_1=M$ to have it valid $\forall k_1\leq M$. In other words, it is sufficient to have $log_2(1+\frac{\beta}{1+\frac{M-1}{M}})+Mlog_2(1-\frac{\rho\delta}{(1+\rho)(1+\delta)})\geq0$ for the objective function to be increasing with $k'_2$.
\end{IEEEproof}
We suppose that the previous condition holds. We can now tackle the effect of adding literals users to our scheduled sets of users.
%
%
\begin{lemma}
If $log_2(1+\frac{\rho}{1+(D-1)\delta})+Dlog_2(1-\frac{\beta}{(\beta+1)(M+1)})\geq0$, then the objective function is increasing with $k_1$ and therefore the optimum is attained for $k_1=M$.
\label{lemma2}
\end{lemma}
\begin{IEEEproof}
As proven in Lemma \ref{lemma1}, the maximum has $k'_2=k_2$ and therefore the objective function is the following:
\begin{equation}
T(k_1,k_2)=\sum_{i=1}^{k_1}log_2(1+\frac{\rho}{1+g_i(k_2)\delta})+\sum_{j=1}^{k_2}log_2(1+\frac{\beta}{1+\frac{f_j(k_1)}{M}})
\end{equation}
To study the effect of increasing the number of literals users $k_1$, we calculate the marginal gain of having an additional literal user scheduled, supposing that it would increase our clauses from $k_2$ to $k_3\geq k_2$:
\begin{equation}
\begin{aligned}
& G=T(k_1+1,k_3)-T(k_1,k_2)=log_2(1+\frac{\rho}{1+g_{k_1+1}(k_3)\delta})+\\
&\sum_{j=1}^{k_3}log_2(1+\frac{\beta}{1+\frac{f_j(k_1+1)}{M}})-\sum_{j=1}^{k_2}log_2(1+\frac{\beta}{1+\frac{f_j(k_1)}{M}})
\end{aligned}
\end{equation}
The worst case for the gain takes place when this additional added literal user add interference on all the clauses users previously scheduled, does not result in an increase of the clauses scheduled and it got the worst case interference possible. In other words, $f_j(k_1+1)=f_j(k_1)+1 \:\: \forall j$, $k_3=k_2$ and $g_{k_1+1}(k_3)=k_2-1$. The lower-bound on the gain becomes:
\begin{equation}
\begin{aligned}
& G\geq log_2(1+\frac{\rho}{1+(k_2-1)\delta}) +\\
&\sum_{j=1}^{k_2}log_2(\frac{\beta M+M+f_j(k_1)+1}{M+f_j(k_1)+1}\frac{M+f_j(k_1)}{\beta M+M+f_j(k_1)})
\end{aligned}
\end{equation}
The second term expression takes the form of $\frac{(a+1)b}{(b+1)a}=1+\frac{b-a}{ab+a}$ with $a=\beta M+M+f_j(k_1)$ and $b=M+f_j(k_1)$. The expression therefore becomes $1-\frac{\beta M}{ab+a}$. By taking into account that $f_j(k_1)\geq0 \:\: \forall j,\forall k_1$, the following inequalities hold : $M\leq b$ and $\beta M+M\leq a$. We can therefore conclude:
\begin{equation}
G\geq log_2(1+\frac{\rho}{1+(k_2-1)\delta})+k_2log_2(1-\frac{\beta}{(\beta+1)(M+1)})
\end{equation}
As previously done, instead of checking $D$ conditions, we construct the following function: $f(x)=log_2(1+\frac{\rho}{1+(x-1)\delta})+xlog_2(1-\frac{\beta}{(\beta+1)(M+1)})$. By deriving with respect to $x$, we have: $f'(x)=\frac{1}{ln(2)}\frac{\frac{-\rho\delta}{(1+(x-1)\delta)^2}}{1+\frac{\rho}{1+(x-1)\delta}}+log_2(1-\frac{\beta}{(\beta+1)(M+1)})$.
One can easily verify that $f'(x)$ is negative $\forall x\geq1$. This tells us that $f(x)$ is actually decreasing with respect to $x$ and therefore if $f(D)\geq0$ for a certain $D\in \mathbb{N}^{*}$, then $f(k_2)\geq0\:\: \forall 
k_2\leq D$. Therefore, it is enough for that condition to be verified for $k_2=D$ to have it valid $\forall k_2\leq D$. In other words, a sufficient condition for the objective function to increase with respect to $k_1$ is that $log_2(1+\frac{\rho}{1+(D-1)\delta})+Dlog_2(1-\frac{\beta}{(\beta+1)(M+1)})\geq0$ which concludes our proof.
\end{IEEEproof}
If we consider that the previous conditions are verified, the maximal rate is achieved when scheduling $M$ literals users:
\begin{equation}
T(M,k_2)=\sum_{i=1}^{M}log_2(1+\frac{\rho}{1+g_i(k_2)\delta})+\sum_{j=1}^{k_2}log_2(1+\frac{\beta}{1+\frac{f_j(M)}{M}})
\label{finalob}
\end{equation}
The last thing to check is the effect of the number of allowed clauses users $k_2$ on the objective function.

\begin{lemma}
If $Mlog_2(\frac{1+\rho+(D-1)\delta}{(1+(D-1)\delta)(1+\rho)})+Dlog_{2}(\frac{\beta M+2M-1}{2M-1+\beta(2M-1)})+log_2(1+\beta)\geq0$, then the objective function increases with the allowed numbers of clauses users. In other words, coupled with the previous conditions, the maximum is achieved for $k_1=M$ while seeking the setting of $M$ literals that leads to the largest number of allowed clauses users $k_2$. 
\label{lemma3}
\end{lemma}
\begin{IEEEproof}
There are many possible M literals configurations, in fact $2^M$ of them, then it is important to discuss which of them leads to the highest objective function.
Suppose we have the setting of $k_1=M$ and $k_2=d_1-1$. Suppose we can have the setting of $k_1=M$ and $k_2=d_1$ and we therefore need to calculate the marginal gain accordingly. The worst marginal gain takes place when we have:
\begin{equation}
\begin{aligned}
& T(M,d_1-1)=Mlog_2(1+\rho)+(d_1-1)log_2(1+\beta)\\
&T(M,d_1)=Mlog_2(1+\frac{\rho}{1+(d_1-1)\delta})+d_1log_{2}(1+\frac{\beta}{1+\frac{M-1}{M}})
\end{aligned}
\end{equation}
This is the case where the $(k_1=M,k_2=d_1-1)$ scenario has no interference at all while for the case of $(k_1=M,k_2=d_1)$, we have the highest interference possible. Therefore a lower-bound on our marginal gain is therefore:
\begin{equation}
\begin{aligned}
& G\geq Mlog_2(\frac{1+\rho+(d_1-1)\delta}{(1+(d_1-1)\delta)(1+\rho)})+\\
&d_1log_{2}(\frac{\beta M+2M-1}{2M-1+\beta(2M-1)})+log_2(1+\beta)
\end{aligned}
\end{equation}
It is enough to suppose that this lower-bound is positive $\forall d_1\leq D$ to prove that our objective function is increasing in the number of \emph{allowed} clauses users. We can further proceed by defining the following function $f(x)=Mlog_2(\frac{1+\rho+(x-1)\delta}{(1+(x-1)\delta)(1+\rho)})+xlog_{2}(\frac{\beta M+2M-1}{2M-1+\beta(2M-1)})+log_2(1+\beta)$.
By deriving with respect to $x$, we have that $f'(x)=\frac{M}{ln(2)}\frac{\frac{-\rho\delta(1+\rho)}{((1+(x-1)\delta)(1+\rho))^2}}{\frac{1+\rho+(x-1)\delta}{(1+(x-1)\delta)(1+\rho)}}+log_{2}(\frac{\beta M+2M-1}{2M-1+\beta(2M-1)})$.
One can easily verify that $f'(x)\leq0 \:\: \forall x\geq1$. This tells us that $f(x)$ is actually decreasing with respect to $x$ and therefore if $f(D)\geq0$ for a certain $D\in \mathbb{N}^{*}$,  then $f(d_1)\geq0\:\: \forall d_1\leq D$.
\end{IEEEproof}
%

\begin{lemma}
For any pair $(M_1,D_1)$, one can easily find an appropriate scenario of our problem where the previous conditions are satisfied.
\label{lemma4}
\end{lemma}
\begin{IEEEproof}
We suppose we have $\delta=\delta_1$. We argue that one can always find $\beta$ and $\rho$ such that all the previous conditions are verified. To proceed with our proof, we first start with condition $1$: the second term $M_1log_2(1-\frac{\rho\delta_1}{(1+\rho)(1+\delta_1)})$ can be lower-bounded by $M_1log_2(1-\frac{\delta_1}{(1+\delta_1)})$ since $\frac{\rho}{\rho+1}<1$. Therefore, it is sufficient to have $\beta\geq((1-\frac{\delta_1}{(1+\delta_1)})^{-M_1}-1)(1+\frac{M_1-1}{M_1})\overset{\Delta}{=}a$ for condition $1$ to be satisfied. For condition $2$, the second term can be lower-bounded by $D_1log_2(1-\frac{1}{M_1+1})$ since $\frac{\beta}{\beta+1}<1$. Hence, it is enough to have $\rho\geq((1-\frac{1}{(M_1+1)})^{-D_1}-1)(1+(D_1-1)\delta_1)\overset{\Delta}{=}b$ to satisfy condition $2$. As for condition $3$, the first term can be written as follows: $M_1log_2(\frac{1+\rho+(D_1-1)\delta_1}{(1+(D-1)\delta_1)(1+\rho)})=M_1log_2(1-\frac{\rho(D_1-1)\delta_1}{(1+(D_1-1)\delta_1)(1+\rho)})$. Since $\frac{\rho}{\rho+1}<1$, this term can be lower-bounded by $M_1log_2(1-\frac{(D_1-1)\delta_1}{(1+(D_1-1)\delta_1)})$. As for the second term, one can easily verify that $\frac{\beta M+2M-1}{2M-1+\beta(2M-1)}\geq\frac{1}{2} \: \forall M\in \mathbb{N}^{*},\beta>0$ and therefore the second term can be lower-bounded by $-D_1$. In this case, it is enough for $\beta$ to verify $\beta\geq2^{D_1}(1-\frac{(D_1-1)\delta_1}{(1+(D-1)\delta_1)})^{-M_1}-1\overset{\Delta}{=}c$ for condition $3$ to be satisfied. To recuperate, it is sufficient to choose $\rho\geq b$ and $\beta\geq$max$\{a,c\}$ for the conditions to verified. We can therefore assert that one can always find a scenario of our problem where all these conditions are verified for any $(M_1,D_1)$ pair chosen which concludes our proof.
\end{IEEEproof}

The results of Lemma \ref{lemma4} are of paramount importance for our proof. It tells us that for any values of $(M_1,D_1)$ (and hence for any mapped SAT problem 
instance), one can always find a particular scenario of our problem where the conditions of Lemmas \ref{lemma1}-\ref{lemma3} are satisfied. Therefore in this case, the objective function is maximized by seeking the setting of $M$ literals that makes $k_2$ as high as possible. We argue that for this 
scenario, one can always find a certain $\gamma$ where the equivalence between our decision problem and the SAT instance takes place. We therefore proceed to writing 
this statement in a rigorous manner. For this purpose, we define $\Phi=\{T(M,k_2): k_2\leq D-1\}$ and let $\gamma_{th}= \displaystyle{\max}\:\: \Phi$. We will use $
\gamma_{th}$ to prove the equivalence 
between our decision problem and the SAT problem instance. Let $\epsilon_3>0$ be a \emph{sufficiently small} positive real number, and define $\gamma=\gamma_{th}+
\epsilon_3$. We argue that for this particular choice of $\gamma$,  the answer to our decision version is equivalent to the SAT instance. In fact, if it is possible to 
achieve\footnote{A sum-rate $\gamma$ is said to be achievable if the sum-rate $\geq\gamma$} an objective function of $\gamma$, the only way to 
achieve it is to have a setting of $M$ literals that allows us to schedule $D$ clauses users. Therefore, the answer to the decision version of our problem for this $
\gamma$ is TRUE \emph{if and only if} our SAT problem is itself TRUE (i.e. all $D$ clauses are scheduled).

Since this analysis is viable for any SAT problem instance (recall the results of Lemma \ref{lemma4}), and knowing that SAT is an NP-complete problem, this proves that the decision 
version of our problem in (\ref{utility}) is NP-hard. In fact, if we are able to solve our decision problem for this $\gamma$ in polynomial time, then we can solve the 
SAT problem in polynomial time as well which is not true, unless P=NP.
This concludes our proof that our problem in (\ref{utility}) is NP-hard.

\end{document}